\documentclass{aa}

\usepackage[square,comma]{natbib}
\usepackage[dvips]{graphics}
\usepackage{epsf}
\usepackage{amssymb}

\newcommand{\IZ}{IZw\,36}
\newcommand{\IZW}{IZw\,18}
\newcommand{\SBS}{SBS\,0335-052}
\newcommand{\Mark}{Markarian\,59}
\newcommand{\1}{\,{\sc i}}
\newcommand{\2}{\,{\sc ii}}
\newcommand{\3}{\,{\sc iii}}
\newcommand{\4}{\,{\sc iv}}
\newcommand{\5}{\,{\sc v}}
\newcommand{\6}{\,{\sc vi}}
\newcommand{\kms}{km\,s$^{-1}$}
\newcommand{\cm}{cm$^{-2}$}

\usepackage{graphicx}
\usepackage{txfonts}
\begin{document}

   \title{Interstellar abundances in the neutral and ionized gas of NGC604}


 \author{V.~Lebouteiller\inst{1}, D.\,Kunth\inst{1}, J.~Lequeux\inst{2},
 A.~Aloisi\inst{3, 4},
J.-M.~D\'esert\inst{1},  G.~H\'ebrard\inst{1}, A.~Lecavelier~des~\'Etangs\inst{1} 
    \and A.~Vidal-Madjar\inst{1}
 }

   \offprints{V. Lebouteiller, \email{leboutei@iap.fr}}

   \institute{1: Institut d'Astrophysique de Paris,
UMR7095 CNRS, Universit{\' e} Pierre \&\ Marie Curie, 98 bis boulevard Arago, 75014 Paris\\
2: LERMA - Observatoire de Paris, 61, Avenue de
l'Observatoire,  F-75014 Paris, France\\
3: Space Telescope Science Institute, 3700 San Martin Drive, Baltimore,
      MD 21218, USA \\
4: On assignment from the Space Telescope Division of ESA }

   \date{Received ; accepted }

  \abstract
   {}
{We present \emph{FUSE} spectra of the giant H\2\ region NGC604 in the spiral galaxy M33.
Chemical abundances are tentatively derived from far-UV absorption lines and compared to
those derived from optical emission lines.}
{Absorption lines from neutral hydrogen and heavy elements were observed against the
continuum provided by the young massive stars embedded in the H\2\ region. We derived the
column densities of H\1, N\1, O\1, Si\2, P\2, Ar\1, and Fe\2, fitting the line profiles
with either a single component or several components. We used \texttt{CLOUDY} to correct
for contamination from the ionized gas. Archival HST/\emph{STIS} spectra across NGC604
allowed us to investigate how inhomogeneities affect the final H\1\ column density.}
{Kinematics show that the neutral gas is physically related to the H\2\ region. The
\emph{STIS} spectra reveal H\1\ column density fluctuations up to 1\,dex across NCG604.
Nevertheless, we find that the H\1\ column density determined from the global
\textit{STIS} spectrum does not differ significantly from the average over the individual
sightlines. Our net results using the column densities derived with \textit{FUSE},
assuming a single component, show that N, O, Si, and Ar are apparently underabundant in
the neutral phase by a factor of $\sim 10$ or more with respect to the ionized phase,
while Fe is the same. However, we discuss the possibility that the absorption lines are
made of individual unresolved components, and find that only P\2, Ar\1, and Fe\2\ lines
should not be affected by the presence of hidden saturated components, while N\1, O\1,
and Si\2 might be much more affected.}
{If N, O, and Si are actually underabundant in the neutral gas of NGC604 with respect to
the ionized gas, this would confirm earlier results obtained for the blue compact dwarfs,
and their interpretations. However, a deeper analysis focused on P, Ar, and Fe mitigates
the above conclusion and indicates that the neutral gas and ionized gas could have
similar abundances.}

\keywords{galaxies: abundances - galaxies: dwarfs - galaxies: ISM
galaxies: starburst ultraviolet: galaxies
 }

\titlerunning{Interstellar abundances in NGC604}
\authorrunning{V. Lebouteiller et al.}

\maketitle

%

\section{Introduction}

The interstellar medium (ISM) is mainly enriched by heavy elements produced by the young
massive stars during many starburst episodes taking place over the star-formation history
of the galaxy. The immediate fate of metals released by these massive stars in the H\2\
regions where stars recently formed, has not yet been settled. Kunth \&\ Sargent (1986)
have suggested that the H\2\ regions of the blue compact dwarf galaxy \IZW\ enrich
themselves with metals expelled by supernov\ae\ and stellar winds during the timescale of
a starburst episode (i.e., a few $10^6$\,yr). However, observational evidence (see e.g.,
Martin et al. 2002) shows that metals might be contained in a hot phase reaching the halo
of galaxies, before they could cool down and eventually mix into the ISM. The issue of
the possible self-enrichment of H\2\ regions is essential to understanding the chemical
evolution of galaxies, since H\2\ region abundances derived from the optical emission
lines of the ionized gas are extensively used to estimate the metallicity of galaxies. If
there is self-pollution of these regions, the derived abundances would no longer reflect
the actual abundances of the ISM.

One approach to studying possible self-enrichment and, more generally, the mixing of
heavy elements in the ISM is to compare the abundances of the ionized gas to those of the
surrounding neutral gas. A first attempt was made by Kunth et al. (1994) using the
\textit{GHRS} onboard the Hubble Space Telescope (HST). These authors derived the neutral
oxygen abundance in \IZW\ using the O\1\ $\lambda$1302 line arising in absorption and
using the H\1\ column density and velocity dispersion from radio observations. They found
an oxygen abundance that is 30 times lower than in the ionized gas. However, the result
remains inconclusive since the O\1\ line is likely to be heavily saturated, so that its
profile could be reproduced with a solar metallicity by choosing a different dispersion
velocity parameter, as pointed out by Pettini \&\ Lipman (1994).

The \textit{Far Ultraviolet Spectroscopic Explorer} (\textit{FUSE}; Moos et al. 2000)
gives access to many transitions of species arising in the neutral gas, including H\1,
N\1, O\1, Si\2, P\2, Ar\1, and Fe\2, with a wide range of oscillator strengths for most
of the species. Hence it becomes possible to determine abundances in the neutral gas with
better accuracy.

A surprise came from recent \textit{FUSE} studies of five gas-rich blue compact
metal-poor galaxies (BCDs): \IZW\ (Aloisi et al. 2003; Lecavelier et al. 2004), \Mark\
(Thuan et al. 2002), \IZ\ (Lebouteiller et al. 2004), \SBS\ (Thuan et al. 2005), and
NGC625 (Cannon et al. 2004). In these galaxies, nitrogen seems to be systematically
underabundant in the neutral gas with respect to the ionized gas of the H\2\ regions. The
oxygen abundance seems either lower in the neutral gas or similar to the one in the
ionized gas. This picture, if true, could offer a new view of the chemical evolution of
the ISM in a galaxy, especially for the metal dispersion and mixing timescales. It could,
however, suffer from uncertainties such as ionization corrections, depletion effects, or
systematic errors due to the multiple sightlines toward individual stars and, possibly,
multiple H\2\ regions within the spectrograph entrance aperture.

In this respect, nearby giant H\2\ regions in spiral galaxies provide an interesting
case, since only one H\2\ region fits the aperture. Moreover, their resolved young
stellar population can be investigated to analyze the stellar continuum and in order to
understand and correct for the effects of having multiple sightlines contributing to the
absorption line profiles. NGC604 is the first nearby giant H\2\ region we have
investigated for this purpose. It is a young star-forming region in the dwarf spiral
galaxy M33 with an age of 3-5\,Myr (e.g., D'Odorico \&\ Rosa 1981; Wilson \&\ Matthews
1995; Pellerin 2006). NGC604 is the brightest extragalactic H\2\ region in the sky after
30~Doradus. It is located about 12' from the center of M33. At a distance of 840\,kpc
(Freedman et al. 1991), 1"~corresponds to 4.1\,pc. The nebula has a core-halo structure,
where the core has an optical diameter of $\sim 220$\,pc and the halo a diameter of $\sim
440$\,pc (Melnick 1980). The optical extinction in NGC604 appears to be correlated with
the brightness. It is highest toward the brightest regions ($A_V = 1.7 \textrm{-}
2.8$\,mag), while the average extinction over the nebula is $A_V \sim 0.5$\,mag,
suggesting that dust is correlated with the ionized gas (Churchwell \&\ Goss 1999,
Viallefond et al. 1992).

\begin{figure}
 \epsfxsize=7cm
\rotatebox{-90}{\epsfbox{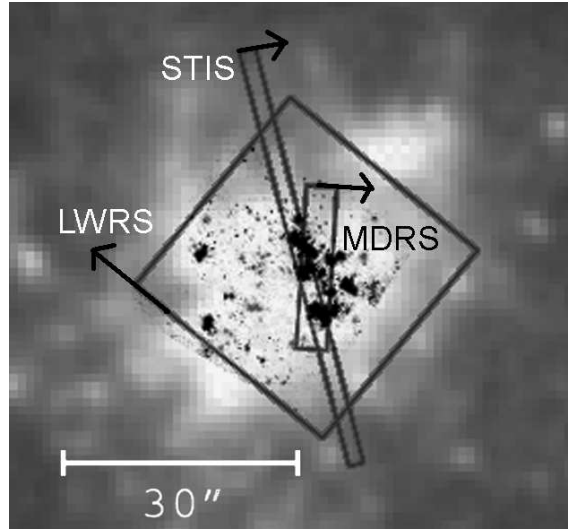}}
  \hspace{0.4cm}
\caption[]{\small{\textit{FUSE} and HST/\textit{STIS} apertures are plotted over the
optical image from the POSS2 survey (B band). We superimposed the most massive stars in
black as observed with \textit{STIS} at 2000\,\AA. The arrows show the dispersion
direction for each slit. North is up.}} \label{fig:slits} \vspace{0.1cm}
\end{figure}

We first describe the observations in Sect.~\ref{sec:obs} and the data analysis in
Sect.~\ref{sec:data}, with particular emphasis on the influence of the source extent on
the absorption line profiles. In Sect.~\ref{sec:mol} we analyze the diffuse molecular
hydrogen content, while in Sect.~\ref{sec:neutral} we infer the neutral hydrogen column
density. Metal column densities in the neutral gas are discussed in
Sect.~\ref{sec:heavy}. We estimate the neutral gas chemical composition in
Sects.~\ref{sec:model} and~\ref{sec:ab} and eventually compare it with the ionized gas of
the H\2\ region. Finally, in Sect.~\ref{sec:hidden}, we investigate how the presence of
hidden saturated unresolved absorption components can be a major problem for determining
the column density.

\section{Observations}\label{sec:obs}

\begin{figure*}
 \epsfxsize=17.5cm
\rotatebox{0}{\epsfbox{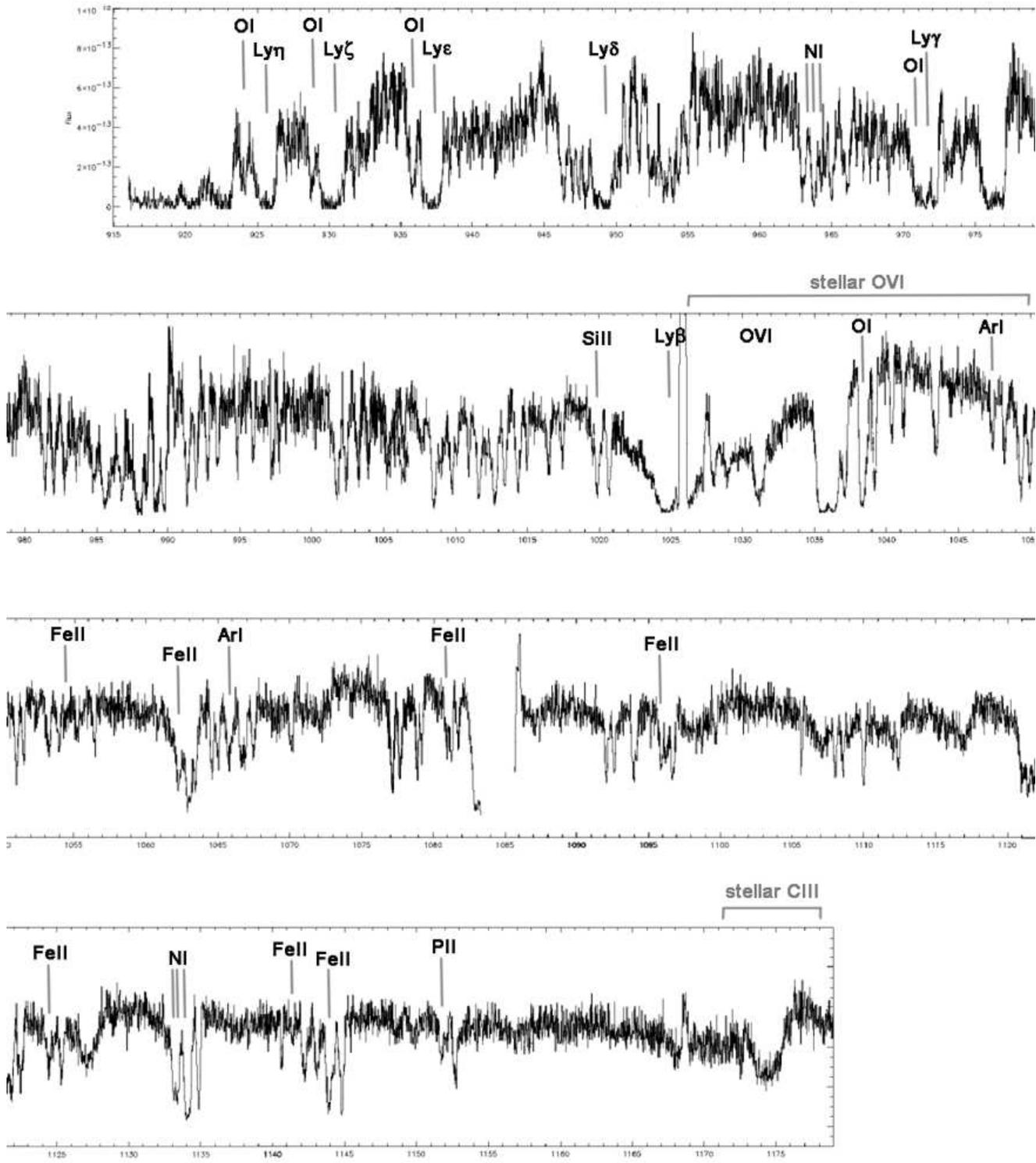}}
  \hspace{0.4cm}
\caption[]{\small{\textit{FUSE} LWRS spectrum of NGC604, indicating the most prominent
interstellar lines from NGC604. We also show the wavelength ranges of the two broad O\6\
and C\3\ stellar P~Cygni profiles.}} \label{fig:spectre} \vspace{0.1cm}
\end{figure*}

The ionizing cluster of NGC604 was observed with the \emph{FUSE} telescope through the
MDRS and LWRS entrances (see the log of the observations in Table~\ref{tab:obs}). The
source as observed in the far-UV, with an apparent size of $\sim$10"$\times$15" (see
Fig.~\ref{fig:slits}), should be considered as extended relative to the size of the
apertures. Data were recorded through the LiF and SiC channels ($\sim 1000 \textrm{-}
1200$\,\AA\ and $\sim 900 \textrm{-} 1100$\,\AA\ resp.). These channels are independently
calibrated and provide redundant data, making it possible to identify possible
instrumental artifacts. The data were processed with the \texttt{CalFuse} $2.4$ pipeline.
Figure~\ref{fig:spectre} shows the LWRS spectrum over the full spectral range $\sim
900\textrm{-}1200$\,\AA. Apart from the lines of neutral species, a remarkable detection
is the interstellar O\6\ line in NGC604, tracing hot gas, at a radial velocity of
$-211\pm25$\,\kms ($\lambda = 1031$ \AA). We did not observe any significant differences
in the extracted spectrum whether correcting for the jitter of the satellite or not.

\begin{table}
\caption{\small{Log of the observations.}}\label{tab:obs}
\begin{center}
\begin{tabular}{lllll}
\hline \hline
   & \textit{FUSE} & \textit{FUSE}  & HST/\textit{STIS}  & \textit{IUE}\\
 \hline
 PID        & B018          & A086              & 9096          &       \\
 Date       & 12/2003       & 09/2001        & 08/1998      &     1979-1984 \\
 Range (\AA)     & 900-1200   & 900-1200   & 1150-1730    &    1150-1950 \\
 Exp. (ksec) & 7         & 13         & 2    &      4.8-22.8  \\
 Aperture   & 4"$\times$20" & 30"$\times$30" & 52"$\times$2" &  $\approx$10"$\times$20" \\
            & (MDRS)        & (LWRS)        &   G140L      &   SWP \\
 Resolution$^{\mathrm{a}}$ & 0.07 \AA  & 0.07 \AA       & 2 \AA &   5 \AA \\
 \hline
\end{tabular}
\end{center}
\begin{list}{}{}
\item[$^{\mathrm{a}}$] Spectral resolution for a point-like source.
\end{list}
\end{table}

We investigated the neutral gas inhomogeneity by using HST/\textit{STIS} spectra toward
individual stars in the ionizing cluster (Sect.~\ref{sec:neutralstis}). These spectra
were taken with the G140L grating using the FUV-MAMA detector, and have been kindly
provided to us by F. Bruhweiler and C. Miskey. The extraction technique is described by
Miskey et al. (2003).

We also used \textit{IUE} archival data in the short-wavelength low-resolution mode to
determine the H\1\ content in NGC604 using the Ly$\alpha$ line
(Sect.~\ref{sec:neutraliue}). We analyzed 10 \textit{IUE} spectra with exposure times
longer than $\sim 5$\,ksec.

\section{Data analysis}\label{sec:data}

The most plausible physical situation in NGC604 is the presence of clouds with different
physical properties that produce different absorption line components. Indeed, the
multiple sightlines toward the massive stars in the NGC604 cluster contribute to the
global spectrum of the region. However, due to resolution effects, we only observe a
single component, and this is the assumption we make in a first analysis presented in
this section. Jenkins et al. (1986) found that this assumption, even in the case of
complex sightlines, does not yield significant systematic errors on the column density
determination, unless very strongly saturated lines are present. We discuss this issue in
detail in Sect.~\ref{sec:hidden}.

\subsection{The profile fitting method}\label{sec:datapfm}

The interstellar absorption lines we observe appear symmetric and can be reproduced by a
Voigt profile. We detect lines arising from both NGC604 and the Milky Way clouds lying
along the sightline. The NGC604 absorption component is blue-shifted by $\sim
-250$\,\kms\ ($\sim 0.8$\,\AA) with respect to the Galactic component ($\sim 0$\,\kms).
Another weaker absorption component is found at $\sim -150$\,\kms, which is likely to be
associated with a high velocity cloud in M33.

The data analysis was performed using the profile fitting procedure \texttt{Owens}
(Lemoine et al. 2002) developed at the Institut d'Astrophysique de Paris by Martin
Lemoine and the \textit{FUSE} French Team. This program returns the most likely values of
many free parameters such as temperatures ($T$), radial velocities ($v$), turbulent
velocity dispersion ($b$), and column densities ($N$) of species, by a $\chi^2$
minimization of the fits of the absorption line profiles. The program also allows changes
in the shape and intensity of the continuum, in the line broadening, and in the zero
level. The errors on the $N$, $b$, $v$, and $T$ parameters are calculated using the
$\Delta \chi^2$ method described in H{\'e}brard et al. (2002), and they include the
uncertainties on all the free parameters (notably the continuum level and shape). All the
errors we report are within $2\,\sigma$.

The \texttt{Owens} procedure is particularly suited to far-UV spectra overcrowded with
absorption lines, since it allows for a simultaneous fit of all lines in a single
spectrum. This method allows us to investigate blended lines with minimum systematic
errors: for example, an Fe\2\ line that is blended with an H$_2$ line can still be used
to constrain the Fe\2\ column density, since the H$_2$ column density is well-constrained
by all the other H$_2$ lines fitted simultaneously in the spectrum.

We derived column densities from profile fitting using two different approaches:
\begin{itemize}
\item \emph{Simultaneous fit}. Species are arranged in several groups, each group defined
by common $b$, $T$, and $v$ parameters. One group refers to species that we assume to be
mainly present in the neutral phase (i.e., H\1, N\1, O\1, Si\2, P\2, Ar\1, and Fe\2),
another group corresponds to the molecular hydrogen H$_2$, another one to the
interstellar O\6, and the last group refers to the other minor species, mainly higher
ionization states.

\item \emph{Independent fits}. Species may not entirely coexist in the same gaseous
phase, hence we no longer assume similar $b$, $T$, and $v$. Each species is defined by
its own physical parameters.
\end{itemize}

We are able to use the NGC604 spectra to compare the two approaches and to discuss in
particular the reliability of the simultaneous-fit method. This one was indeed preferred
for spectra of blue compact dwarf galaxies (see the references in the introduction) $-$
generally because of a relatively low signal-to-noise ratio $-$ but could introduce
systematic errors with respect to the more realistic independent-fit method.

\subsection{Stellar contamination and continuum}\label{sec:datacont}

In our observations, the continuum arises from the combination of the spectra of numerous
UV bright stars.

\begin{figure*}
 \epsfxsize=8.8cm
\rotatebox{0}{\epsfbox{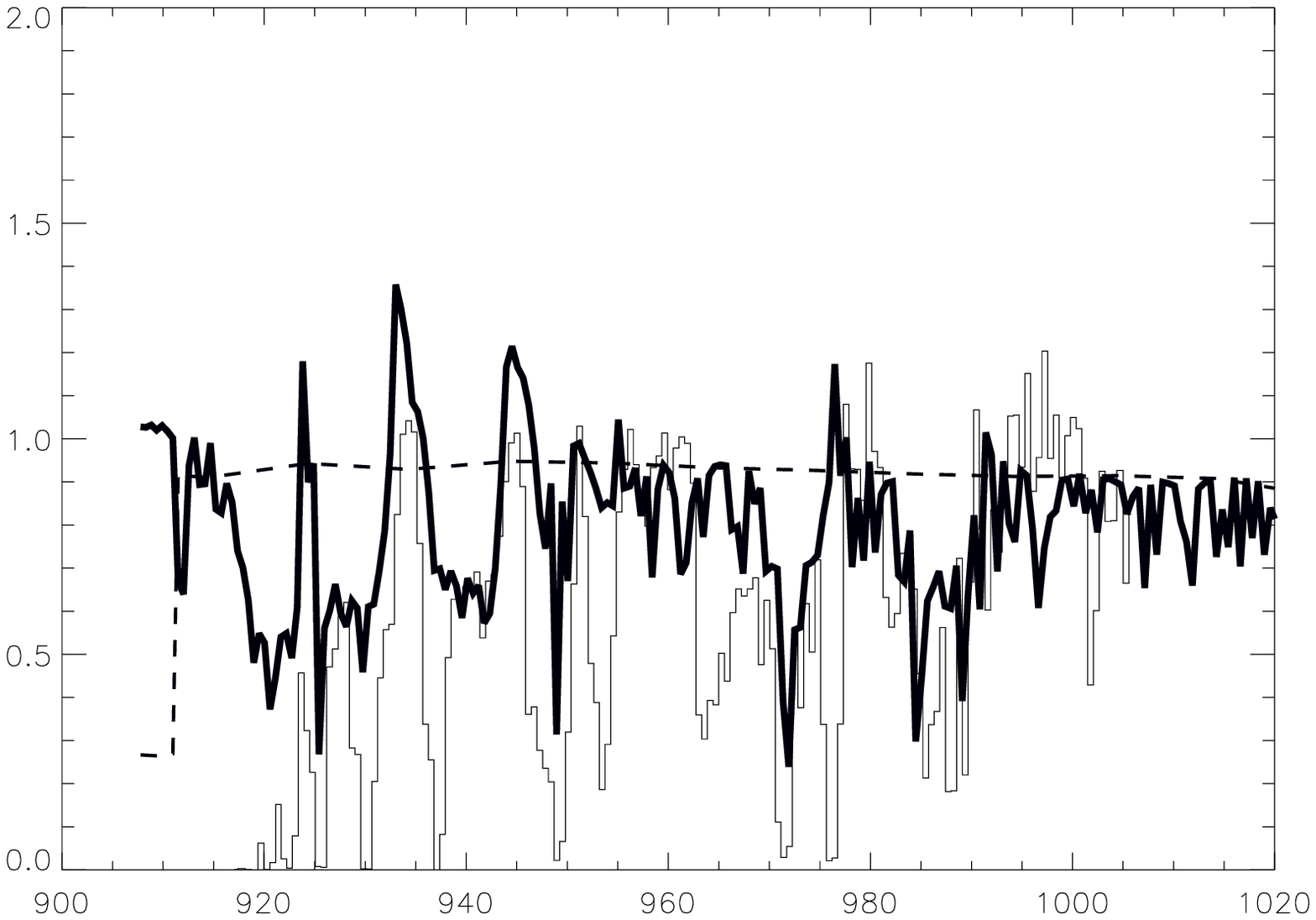}}
 \epsfxsize=8.8cm
\rotatebox{0}{\epsfbox{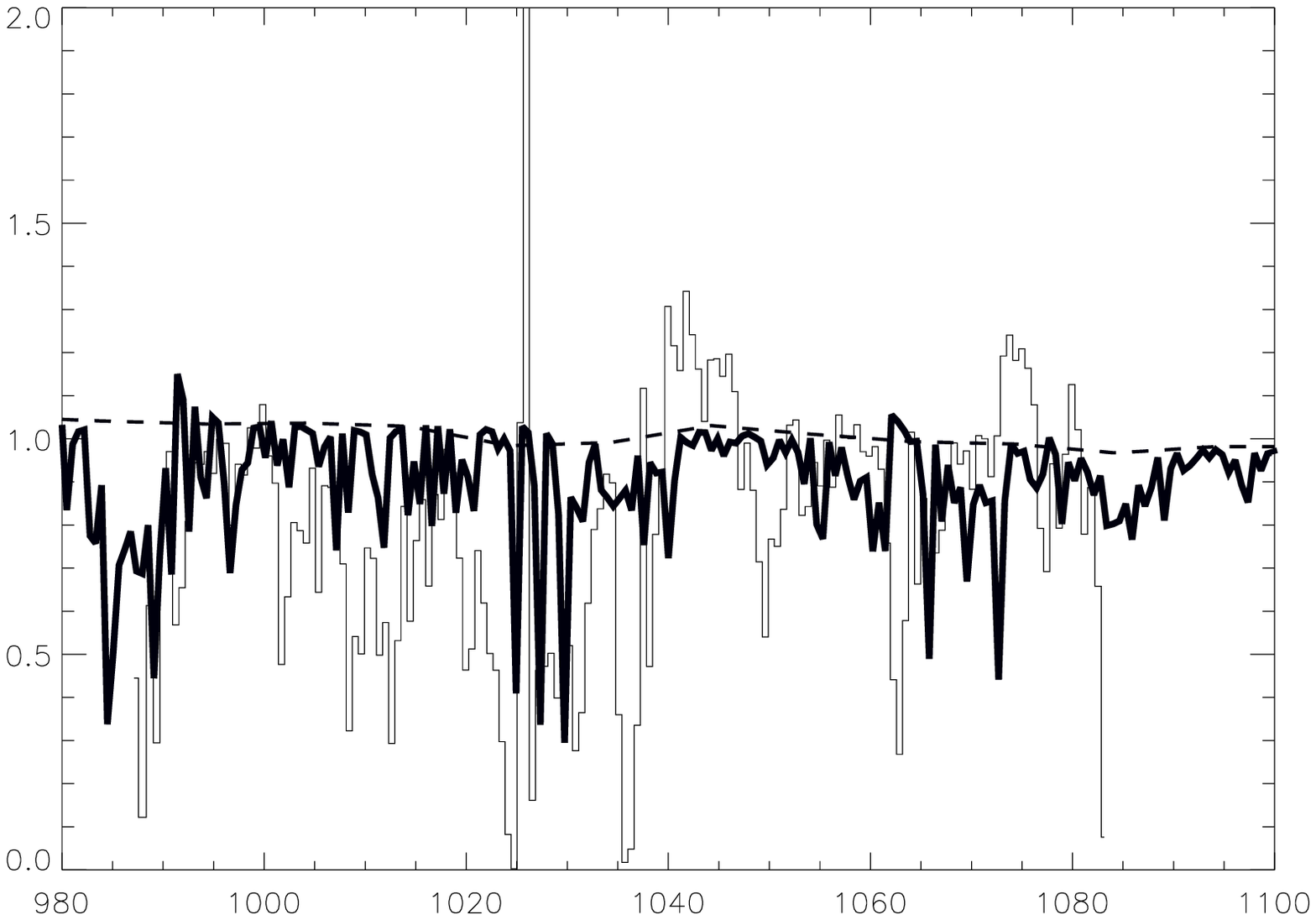}}
 \epsfxsize=8.8cm
\rotatebox{0}{\epsfbox{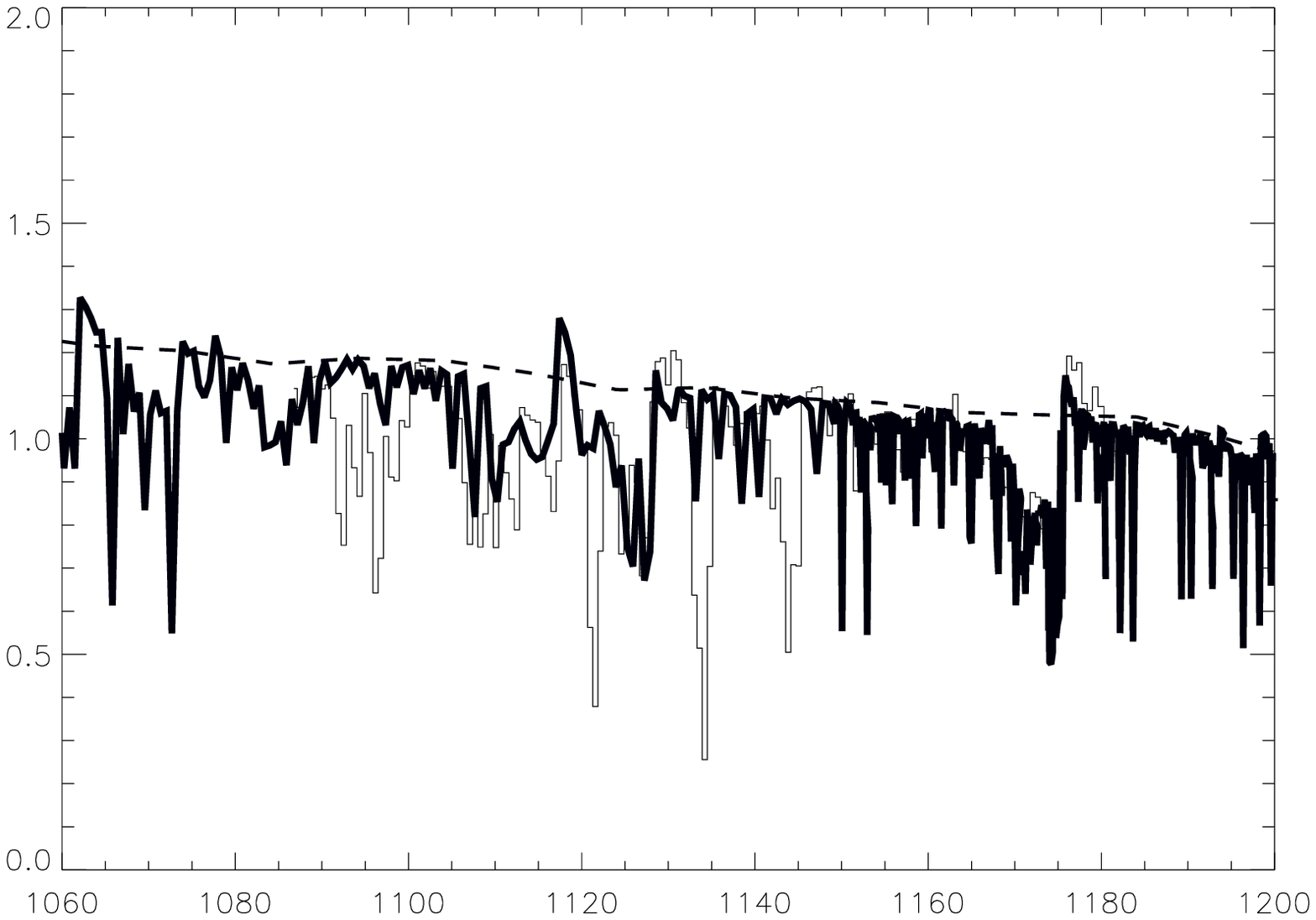}}
 \epsfxsize=8.8cm
\rotatebox{0}{\epsfbox{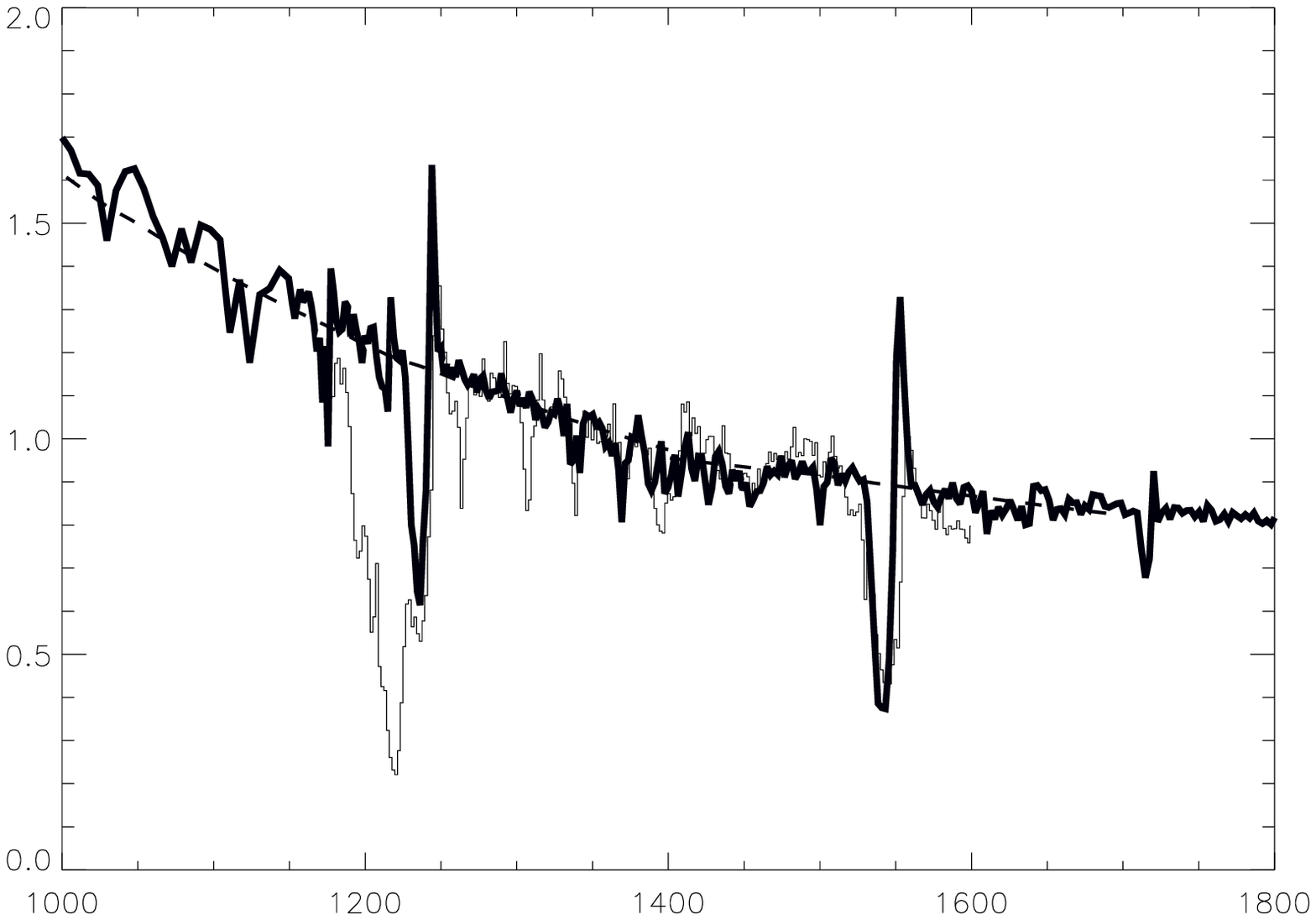}} \caption[]{\small{Theoretical stellar spectra (thick
solid lines) are overplotted onto the observed \textit{FUSE} and HST/\textit{STIS} data
(histograms). The flux (in arbitrary units) is plotted versus the wavelength (\AA) for
the three spectral regions between 900\,\AA\ and 1200\,\AA\ (\textit{FUSE} wavelength
range), and for the spectral window 1200-1800\,\AA\ (\textit{STIS} wavelength range).}}
\label{fig:stellspec} \vspace{0.1cm}
\end{figure*}

Contamination by stellar absorption lines could be an issue for the interstellar line
fitting, especially for H\1. However, the contribution of the photospheric H\1\ lines
becomes significant only when early B stars dominate (Gonzalez et al. 1997), i.e., for
bursts older than $\sim 10$\,Myr (Robert et al. 2003). The age of the burst in NGC604 is
$\sim$3-5\,Myr (see introduction). Hence, given this relatively young age, we expect the
photospheric H\1\ lines to be relatively narrow, and their contribution to be negligible
as compared to the interstellar damped profile. We verified this assumption by comparing
the observed {\it FUSE} and HST/\textit{STIS} spectra with an \textit{ad hoc}
Starburst~99 synthetic model of a young stellar population. The latter was kindly
provided by C. Leitherer and F. Bresolin (private communication), and it makes use of a
new library of theoretical stellar atmospheres that does not include interstellar
absorption lines. The adopted spectra assume an instantaneous 3.5\,Myr old burst and a
gas metallicity of 0.4\,Z$_\odot$. The interstellar H\1\ lines (see for instance
Ly$\beta$ at 1025.7\,\AA\ and Ly$\alpha$ at 1215.7\,\AA) are not significantly
contaminated by the $-$ relatively narrow $-$ photospheric lines (see
Fig.~\ref{fig:stellspec}). Notice that the region around the O\6\ ($\sim 1040$ \AA)
stellar line is not reproduced well because the input physics is still missing from the
theoretical stellar atmosphere models we use.

This synthetic spectral model was also used as a safety check to compare with $-$ but not
to constrain $-$ the continuum we adopted for the line profile fitting with
\texttt{Owens} (see next sections), and no significant differences were found.

\subsection{Influence of the apparent extent of the UV cluster on the line profiles}\label{sec:ext}

Interstellar absorption lines are broader in the \textit{FUSE} LWRS spectrum than in the
MDRS spectrum. The line broadening results from the convolution of several effects:
\begin{itemize}

\item The intrinsic line width, due to the thermal and the turbulent
velocities. Considering the large turbulent velocities usually
measured in BCDs and in NGC604 (see Sect.~\ref{sec:cine}), the thermal
component should be negligible:\\
$\Delta \nu_D = \frac{\nu_0}{c} \times \sqrt{\frac{2kT}{m} + b^{2}}
\approx \frac{\nu_0}{c} \times b$,\\
where $\nu_0$ is the frequency and $m$ the mass of the species under consideration.

\item The radial velocity dispersion of the clouds within the aperture. This effect is
expected to be weaker for a single H\2\ region than for observations of entire galaxies
containing several H\2\ regions.

\item The instrumental line spread function (LSF). For bright point-like sources observed
with \textit{FUSE}, the FWHM of the LSF is $\sim 12$~pixels, i.e., $\sim 20$\,\kms\
(H{\'e}brard et al. 2002).

\item The misalignments of the individual sub-exposures. The final spectrum results from
the co-addition of several exposures with possible wavelength shifts between individual
spectra. This unavoidably introduces some misalignments because of the relatively low
signa-to-noise ratio of each exposure. However, given the relatively good quality of the
NGC604 observations, the misalignments do not yield a significant additional broadening.

\item Finally, the spatial distribution of the UV-bright stars along the slit dispersion
axis that results in wavelength smearing.

\end{itemize}

When comparing the \emph{FUSE} spectra with the two different apertures, only the
geometric effect due to the distribution of the massive stars within the apertures is
likely to vary significantly. The \texttt{Owens} procedure makes it possible to estimate
the line broadening parameter, which accounts for all the effects mentioned above, except
the intrinsic line width, implying the turbulent velocity and the temperature, which is
treated independently. Hence, we have here the opportunity of quantifying in a first
approximation the aperture effects related to the observation of an extended source.

The most likely value of the line-broadening parameter for the MDRS (4"$\times$20"
aperture) spectrum is $\sim 15$ pixels (i.e., $\sim 25$\,\kms), which is close to the
point-source LSF. By taking into account that a 30" source spreads the line by 100 \kms\
(60~pixels), this translates into a spatial extent of about 5", which is comparable to
the size of the MDRS aperture in the dispersion direction. The most likely broadening
value for the LWRS (30"$\times$30") data is instead about 30 pixels. This is equivalent
to a spatial extent of 16", which is smaller than the size of the aperture and therefore
would correspond to the actual extent of the UV bright stellar cluster (see the
HST/\textit{STIS} image at 2000\,\AA\ for comparison in Fig.~\ref{fig:slits}).

As a conclusion, the line broadening parameter we infer for the {\it FUSE} observations
is consistent with the extent of the source at the observed wavelength as constrained by
the aperture size. The column densities derived in Sect.~\ref{sec:metals} take this
broadening effect into account.

\section{Molecular hydrogen}\label{sec:mol}

\begin{table*}
\caption{\small{H$_2$ column densities derived from the two \textit{FUSE} observations.
Errors are at $2 \sigma$.}}\label{tab:CDmol}
\begin{center}
\begin{tabular}{lllllll}
\hline \hline
     & H$_{\rm{2, J=0}}$ & H$_{\rm{2, J=1}}$ & H$_{\rm{2, J=2}}$ & H$_{\rm{2, J=3}}$ & H$_{\rm{2, J=4}}$ & H$_{\rm{2,tot}}^{\mathrm{a}}$ \\
\hline
LWRS & $16.39^{+0.20}_{-0.29}$ & $16.51^{+0.23}_{-0.31}$ & $15.12^{+0.49}_{-0.53}$ & $16.17^{+0.30}_{-0.57}$  & $14.40^{+0.48}_{-0.60}$ & $\textbf{16.86}^{+0.25}_{-0.34}$\\
detection level$^{\mathrm{b}}$ & $\sim 12\sigma$ & $\sim 14\sigma$  & $\sim 7\sigma$ &  $\sim 13\sigma$ & $\sim 4\sigma$ &  \\
MDRS & $16.08^{+0.32}_{-0.47}$ & $16.60^{+0.16}_{-0.28}$ & $15.08^{+0.58}_{-1.20}$ & $16.29^{+0.22}_{-0.39}$  & $14.30^{+1.12}_{-3.30}$ & $\textbf{16.86}^{+0.23}_{-0.36}$ \\
detection level$^{\mathrm{b}}$ &  $\sim 6\sigma$ &  $\sim 6\sigma$ & $\sim 2.5\sigma$  & $\sim 6\sigma$   & $\sim 2\sigma$  &   \\
 \hline
\end{tabular}
\end{center}
\begin{list}{}{}
\item[$^{\mathrm{a}}$] Value obtained by summing over all detected rotational states.
\item[$^{\mathrm{b}}$] Calculated with the $\Delta \chi^2$ method, using all the observed
lines.
\end{list}
\end{table*}

\textit{FUSE} gives access to many molecular hydrogen lines of the Lyman and Werner
bands. The H$_2$ lines arising from Galactic clouds along the sightline are clearly
present in our spectra at the radial velocity of $4\pm5$\,\kms. These lines are
responsible for blending other atomic lines, including lines of the neutral species in
NGC604. However, the column density of each rotational state of the Galactic H$_2$ is
well-constrained given the large number of lines available, allowing correction for this
blending effect.

\begin{figure}
 \epsfxsize=7.0cm
\rotatebox{-90}{\epsfbox{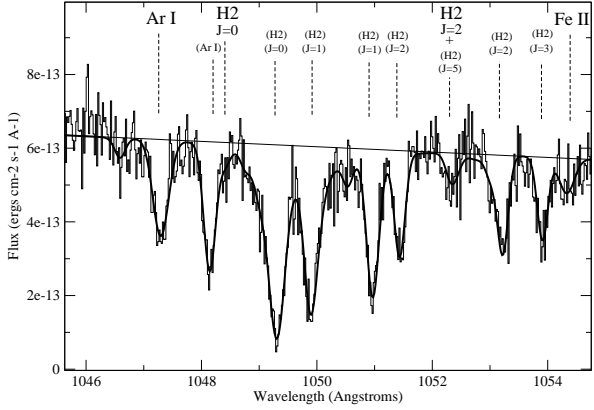}} \hspace{0.4cm} \caption[]{Detection of molecular
hydrogen lines (at $\sim 1048.4$\,\AA\ and $\sim 1052.3$\,\AA) arising in NGC604. The
thick line represents the fit to the data (histogram, {FUSE} MDRS spectrum). Lines
labelled between parentheses refer to the Milky Way component, the others arise in
NGC604. } \label{fig:H2detect} \vspace{0.1cm}
\end{figure}

In contrast to Bluhm et al. (2003), we detected H$_2$ in NGC604 in both LWRS and MDRS
spectra (see Fig.~\ref{fig:H2detect}). This positive detection has been enhanced by the
opportunity given by the \texttt{Owens} procedure to fit simultaneously $-$ and thus
gathering all the information of $-$ all the H$_2$ absorption lines.

\begin{figure}
 \epsfxsize=8cm
\rotatebox{0}{\epsfbox{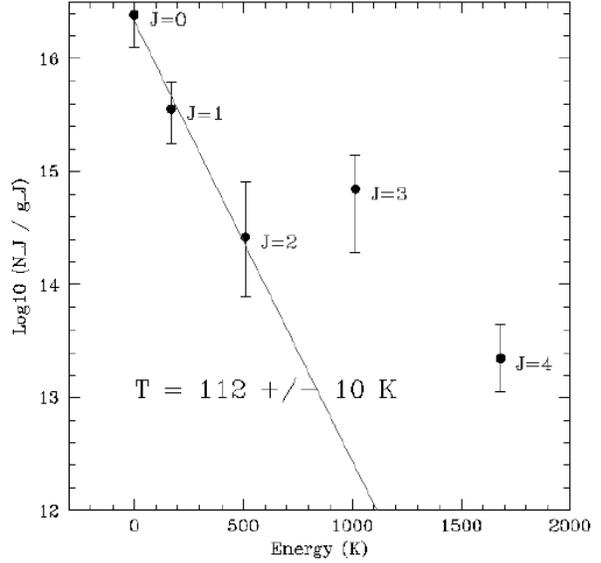}}
  \hspace{0.4cm}
\caption[]{\small{Excitation diagram of the diffuse H$_2$ in NGC604. Column densities of
the H$_2$ rotational levels divided by their statistical weights are plotted against their
excitation energies. The ratio of $J=1$ to $J=0$ levels yields a rotational temperature
$T=112\pm10$\,K.}}
 \label{fig:H2plot} \vspace{0.1cm}
\end{figure}

The velocity we infer for H$_2$ lines in NGC604 is $- 250.4\pm2.1$\,\kms\ in the LWRS
observation and $- 252.2\pm3.4$\,\kms\ in the MDRS. We did not detect H$_2$ from the weak
absorption component at $\sim -150$\,\kms. In Table~\ref{tab:CDmol} we report the
detection levels and the column densities of each rotational state. We used these column
densities to build the excitation diagram of Fig.~\ref{fig:H2plot}. The ratio between the
H$_2$ column densities in the $J=1$ and $J=0$ levels (ortho- to para-hydrogen) yields a
rotational temperature of $T=112 \pm 10$\,K. This temperature can be identified as the
gas kinetic temperature, since these two rotational states should be populated mainly by
collisional processes. The higher excitation of levels $J=3$ and $J=4$ is common in the
ISM and is due to non-collisional processes, such as UV photon pumping, shocks, and
formation of H$_2$ on dust grains.

Given the fact that upper rotational states ($J>4$) are not detected, we choose to
neglect their contribution to calculate the total molecular hydrogen column density which
is found to be $\log N(\rm{H}_2)=16.86^{+0.25}_{-0.34}$. The molecular fraction defined
as $f_{\rm{H2}} = 2 \times N(\textrm{H}_2) / [2 \times N(\textrm{H}_2) +
N(\textrm{H\1})]$ is then $2.6 \times 10^{-4}$ (see Sect.~\ref{sec:neutral} for the
adopted value of the H\1\ column density). Such a low value can seem somewhat surprising
in a star-forming region considering that H$_2$ is an important gas reservoir in which to
form stars. Hoopes et al. (2004) found similarly low fractions in their sample of
starburst galaxies, where upper limits of $f_{\rm{H2}}$ range from $4.9 \times 10^{-6}$
to $1.6 \times 10^{-4}$. It is likely that the diffuse molecular hydrogen is destroyed by
the incident UV flux from the massive stars in the star-forming region. Most of the
remaining molecules should be in dense clumps, which are opaque to far-UV radiation, and
do not contribute to the observed spectra (Vidal-Madjar et al. 2000; Hoopes et al. 2004).

\section{Neutral hydrogen toward NGC604}\label{sec:neutral}

The broad interstellar H\1\ absorption measured toward NGC604 cluster results from the
blended lines arising in the Milky Way, in NGC604, and in the weak component at $\sim -
150$\,\kms. Note that, given the intermediate velocity of the latter, its H\1\ absorption
falls in the heavily saturated core of the main H\1\ absorption line. Therefore, this
component does not significantly modify the integrated H\1\ absorption line profile along
the sightline.

\subsection{Galactic component}\label{sec:galactic}

Interstellar clouds of the Milky Way contribute to the absorption line spectra toward
NGC604. However, according to Velden (1970), the Galactic H\1\ column density is expected
to be somewhat lower than the intrinsic H\1\ component. From the survey of Heiles (1975),
the Galactic H\1\ column density should be a few $10^{20}$ \cm. This is consistent with
the estimates we obtained from the reddening of E(B-V)=0.09 (Israel \&\ Kennicutt 1980).
Indeed, by assuming that $N_{\rm{HI}}/\textrm{E(B-V)} = 5.8 \times 10^{21}$ \cm\
mag$^{-1}$ (Bohlin et al. 1978), we find a Galactic H\1\ column density of $\log
N\textrm{(H\1)} \approx 20.7$ (where the column density is expressed in atom \cm). This
value agrees well with our estimate obtained from the FUSE data (see
Sect.~\ref{sec:neutralfuse}).

\subsection{\textit{FUSE} observations of Ly$\beta$}\label{sec:neutralfuse}

The H\1\ Lyman lines from Ly$\beta$ to Ly$\mu$ fall within the \textit{FUSE} wavelength
range. However, in order to infer the H\1\ column density from \textit{FUSE} spectra, we
can only reliably use Ly$\beta$, since it is the only H\1\ line showing damping wings.
The other Lyman lines fall in the flat part of the curve of growth, where there is
degeneracy between the turbulent velocity and the column density. In addition, the
higher-order H\1\ Lyman lines are located in spectral regions that are more crowded by
other contaminating absorption lines (e.g., Galactic H$_2$ or other atomic interstellar
lines).

\begin{figure}
\epsfxsize=7cm \rotatebox{-90}{\epsfbox{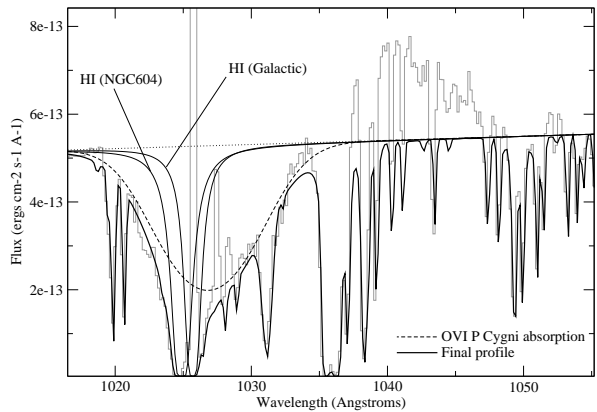}}
  \hspace{0.2cm}
\caption[]{\small{The interstellar Ly$\beta$ profile as observed in the \textit{FUSE}
LWRS spectrum is heavily contaminated by the stellar O\6\ P~Cygni doublet absorption. The
emission component of the latter was not accounted for in the fitting (see text). For
display purposes, all the other interstellar absorption lines are not marked, and data
are rebinned by a factor 8.}} \label{fig:hiovi} \vspace{0.1cm}
\end{figure}

Nevertheless, the H\1\ Lyman~$\beta$ line in NGC604 is contaminated by the stellar O\6\
P~Cygni doublet at 1031.9\,\AA\ and 1037.6\,\AA, as also observed by Cannon et al. (2004)
in the starburst galaxy NGC625. The two stellar lines are heavily blended, resulting in a
broad P~Cygni shaped line. In an attempt to model the stellar O\6\ feature in order to
further estimate the interstellar H\1\ absorption, we decided to reproduce the absorption
component of the total P~Cygni profile using a Gaussian profile. This is a rough
approximation since the theoretical asymmetrical P~Cygni profile should be used instead.
However, one can reasonably expect the global profile, which is the combination of the
individual profiles of different types of stars, to have a roughly symmetrical absorption
component (the C\3, C\4, and Si\4 stellar absorptions indeed appear symmetrical in the
\textit{FUSE} and \textit{IUE} spectra). We performed a simultaneous fit of the
interstellar lines, in particular the H\1\ line Ly$\beta$ in NGC604, together with the
broad O\6\ stellar absorption. The velocity shift and the width of the latter were
allowed to vary without any constraints. The emission component of the P~Cygni profile
was ignored for the fitting.

\begin{figure*}
\epsfxsize=12cm \rotatebox{-90}{\epsfbox{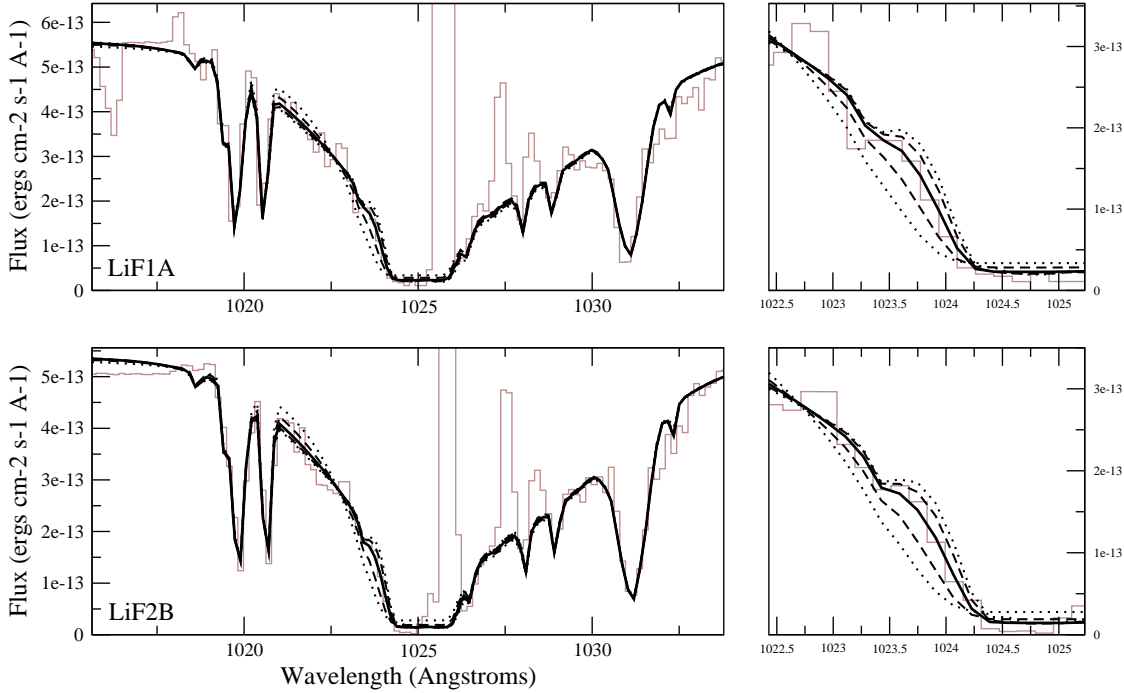}}
  \hspace{0.2cm}
\caption[]{\small{Profiles for an H\1\ column density in NGC604 of $\log N\textrm{(H\1)}
= 20.75$ (thick line), $\log N\textrm{(H\1)} = 20.75\pm0.30$ (dashed lines), and $\log
N\textrm{(H\1)} = 20.75\pm0.60$ (dotted lines). Column densities of other species
(including Galactic H\1) are considered as free parameters in the fitting routine.}}
\label{fig:graphHIref} \vspace{0.1cm}
\end{figure*}

The best result, corresponding to the minimum $\chi^2$, is shown in Fig.~\ref{fig:hiovi}.
The stellar absorption has a likely width of $\sim 1300$\,\kms\ and a blue shift of $\sim
-1500$\,\kms. An H\1\ column density of $\log N\textrm{(H\1)} = 20.7$ was derived for the
Milky Way component (estimated uncertainty 0.3~dex at 2~$\sigma$), while a value of $\log
N\textrm{(H\1)} = 20.75$ ($\pm0.3$, see Fig.~\ref{fig:graphHIref}) was inferred for the
H\1\ in NGC604.

\subsection{\textit{IUE} observations of Ly$\alpha$}\label{sec:neutraliue}

The H\1\ column density can also be determined from the Ly$\alpha$ absorption signature
in low-resolution SWP-\textit{IUE} spectra (see Table~\ref{tab:obs}). We investigated
Ly$\alpha$ from 10 independent spectra (Table~\ref{tab:hiiuetab}).

\begin{figure}
 \epsfxsize=7cm
\rotatebox{-90}{\epsfbox{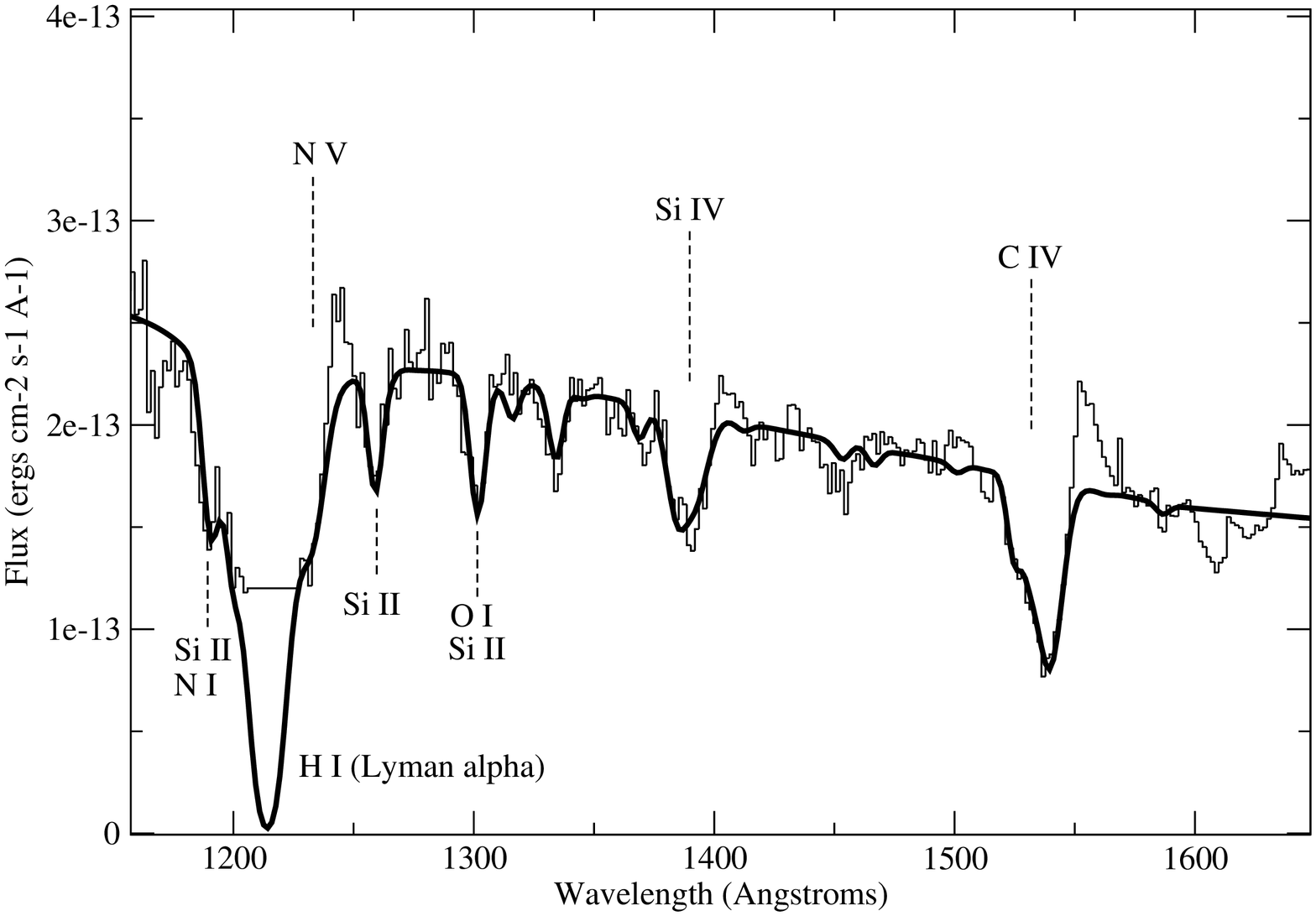}}
  \hspace{0.4cm}
\caption[]{\small{Example of a \textit{IUE} spectrum toward NGC604. The thick line
represents the fit to the data (histogram). Lines marked upwards are stellar in origin.
The other lines are interstellar. The airglow emission around 1210\,\AA\ has been
removed. The emission components of the P~Cygni profiles have not been taken into account
in the profile fitting.}} \label{fig:hiiue} \vspace{0.1cm}
\end{figure}

For the profile fitting, we used the Galactic H\1\ value of $\log N\textrm{(H\1)} =
20.7$, which is our best guess from the \textit{FUSE} observations. In each spectrum,
Ly$\alpha$ is blended with a single stellar N\5\ P~Cygni profile (see
Fig.~\ref{fig:hiiue}). To estimate this contamination, we decided to model the stellar
N\5\ absorption component using the same technique as in the previous section. The mean
width of the stellar absorption N\5\ in the \textit{IUE} spectra was found to be $\sim
1200\pm200$\,\kms\ and the mean velocity shift $\sim -1600\pm150$\,\kms. As a result, we
inferred a mean H\1\ column density in NGC604 of $\log N\textrm{(H\1)} =
21.07^{+0.17}_{-0.24}$.

\begin{table}
\caption{\small{H\1\ column density determinations (in \cm, logarithmic units) derived
from the \textit{IUE} spectra of NGC604.}}\label{tab:hiiuetab}
\begin{center}
\begin{tabular}{lcc}
\hline \hline
Dataset     & Exp. Time (ksec) & $\log N$(H\1) \\
\hline
swp16034 & 10.8 & $20.99^{+0.27}_{-0.72}$ \\
swp16035 & 11.3 & $21.17^{+0.12}_{-0.19}$ \\
swp19154 & 18.0 & $< 21.30^\textrm{a}$ \\
swp19181 & 15.3 & $21.02^{+0.16}_{-0.20}$ \\
swp24509 & 15.6 & $20.85^{+0.11}_{-0.14}$ \\
swp04162 & 4.8  & $21.21^{+0.22}_{-0.40}$ \\
swp05682 & 6.0  & $21.01^{+0.29}_{-0.68}$ \\
swp06638 & 4.8  & $< 21.33^{\mathrm{a}}$ \\
swp07349 & 6.4  & $21.27^{+0.07}_{-0.05}$ \\
swp24508 & 5.4  & $20.87^{+0.13}_{-0.09}$ \\
   \hline
 \multicolumn{2}{c}{Mean column density}  & $21.07^{+0.17}_{-0.24}$  \\
   \hline
\end{tabular}
\end{center}
\begin{list}{}{}
\item[$^{\mathrm{a}}$] The H\1\ Ly$\alpha$ absorption is severely contaminated by a
terrestrial airglow.
\end{list}
\end{table}

\subsection{H\1\ inhomogeneities revealed by \textit{STIS}}\label{sec:neutralstis}

The H\1\ column density derived from \textit{FUSE} and \textit{IUE} spectra could be
different from the true H\1\ column density, due to the presence of multiple unresolved
absorbers along the many contributing sightlines. We investigated this issue and inquired
how the result is affected by the inhomogeneities in the H\1\ by analyzing individual
stellar spectra from the long-slit, spatially resolved, HST/\textit{STIS} data. A few
sightlines were impossible to investigate because of edge effects in the MAMA detector,
together with a location of the star at the border of the slit.

We first performed a profile fitting of Ly$\alpha$ in the single stellar spectra in order
to infer the column density of the H\1\ for each sightline. We adjusted the profile of
the stellar N\5\ line using the method described in the previous sections. We also
assumed that the Galactic H\1\ column density is identical for all the sightlines, given
the relatively low angular extent of NGC604. The results of the profile fitting are
reported in Table~\ref{tab:hististab}. We find variations up to 1~dex in the H\1\ column
density of NGC604, as compared to the uncertainties on the order of $\lesssim 0.4$~dex,
suggesting inhomogeneities of the diffuse neutral gas in front of the ionizing cluster.
This could be a source of systematic errors when determining the total column density
from the integrated light of the cluster. However in our case, the mean value of the H\1\
column density over each single sightline is comparable to the mean value weighted by the
star luminosity (see Table~\ref{tab:hististab}), which is what we actually measure in the
global spectrum of a star cluster.

\begin{table}
\caption{\small{H\1\ column density toward individual stars of NGC604 cluster, as derived
from the Ly$\alpha$ line detected with HST/\textit{STIS}.}}\label{tab:hististab}
\begin{center}
\begin{tabular}{llcc}
\hline \hline
 Star$^{\mathrm{a}}$ & Spectral type$^{\mathrm{a}}$ & Flux at $\sim$1280 \AA$^{\mathrm{b}}$ & $\log N$(H\1) \\
\hline
117 & O4 II & 4.0 & $20.26^{+0.29}_{-0.40}$ \\%
564 & O9 II & 2.8 & $20.95^{+0.16}_{-0.19}$ \\%
578b & O9 Ia & 4.5 & $20.87^{+0.05}_{-0.16}$  \\%
675 & O7 II & 2.8 & $21.33^{+0.16}_{-0.13}$ \\%
690a & O5 III & 1.8 & $20.78^{+0.36}_{-0.75}$ \\%
690b & B0 Ib & 3.4 &  $20.62^{+0.26}_{-0.40}$ \\%
825 & O5 II & 3.5 & $20.36^{+0.36}_{-1.32}$ \\%
867a & O4 Iab & 10.0 & $20.70^{+0.25}_{-0.44}$  \\%
867b & O4 Ia & 22.0 & $20.71^{+0.29}_{-0.49}$  \\%
\hline
\multicolumn{3}{c}{Mean column density}  & $20.84^{+0.20}_{-0.26}$   \\
\multicolumn{3}{c}{Flux-weighted mean column density}   & $20.77^{+0.23}_{-0.33}$   \\
   \hline
\end{tabular}
\end{center}
\begin{list}{}{}
\item[$^{\mathrm{a}}$] From Miskey et al. (2003). \item[$^{\mathrm{b}}$] In units of
$\times10^{-15}$ ergs \cm\ s$^{-1}$ \AA$^{-1}$.
\end{list}
\end{table}

\begin{figure}
\epsfxsize=7cm \rotatebox{-90}{\epsfbox{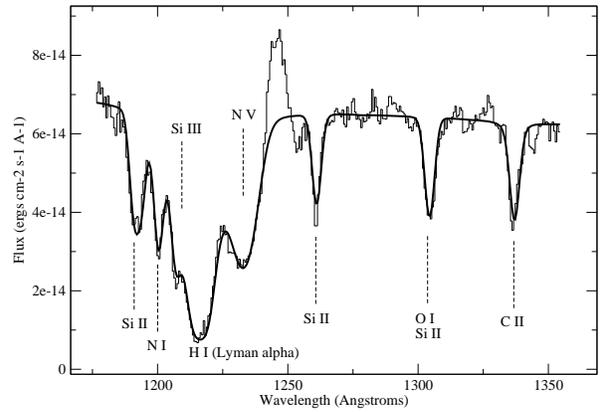}}
  \hspace{0.2cm}
\caption[]{\small{Profile fitting of the Ly$\alpha$ line from the global
HST/\textit{STIS} spectrum (obtained by summing the individual extracted stellar
spectra). See Fig.~\ref{fig:hiiue} for a description of the plot.}} \label{fig:alpha}
\vspace{0.1cm}
\end{figure}

In addition, we synthesized the global spectrum within the \textit{STIS} 52"$\times$2"
slit, summing each one of the extracted stellar spectra. In this way, we simulated the
spectral observation of a cluster as a whole, i.e., our \textit{FUSE} observations. Given
the relatively high dispersion of the G140L grating ($\sim 2$\,\AA), combined with the
width of the \textit{STIS} slit, different positions of stars within the slit can lead to
in a relative wavelength shift up to several thousands of \kms. We thus corrected the
data for these shifts. This allowed us to reach an accuracy on the order of 1~pixel,
corresponding to a velocity dispersion of less than $\sim 150$\,\kms, comparable to the
wavelength smearing along the \textit{FUSE} LWRS aperture ($\sim 50$\,\kms,
Sect.~\ref{sec:ext}). From the Ly$\alpha$ profile of this summed spectrum (see
Fig.~\ref{fig:alpha}), we obtained a column density of $\log N\textrm{(H\1)} =
20.67^{+0.19}_{-0.22}$ for the H\1\ in NGC604. This estimate is consistent within the
errors with the "simple" and "luminosity-weighted" means reported in
Table~\ref{tab:hististab}. This indicates that, in our integrated spectra, we are not
misestimating the actual column density, even though the absorption arises from many
sightlines intersecting clouds with different physical properties.

\subsection{The adopted H\1\ column density in NGC604}\label{sec:neutraladopted}

\begin{table*}
\caption{\small{H\1\ column density determinations using \textit{FUSE}, \textit{IUE} and
HST/\textit{STIS} data.}}\label{tab:hidet}
\begin{center}
\begin{tabular}{llll}
\hline \hline
Instrument & Line & Remarks & Value \\
\hline
  \textit{FUSE} & Ly$\beta$ & Corrected for the stellar O\6\ contamination & $20.75 \pm 0.30$ \\
 \textit{IUE} & Ly$\alpha$ &Mean value of the sample spectra & $21.07^{+0.17}_{-0.24}$ \\
 HST/\textit{STIS} & Ly$\alpha$ &Mean value from the individual star spectra&  $20.84^{+0.20}_{-0.26}$ \\
 HST/\textit{STIS} & Ly$\alpha$ &Flux-weighted mean value from the individual star spectra &  $20.77^{+0.23}_{-0.33}$ \\
 HST/\textit{STIS} & Ly$\alpha$ &Profile fitting of the global spectrum & $20.67^{+0.19}_{-0.22}$ \\
 Radio & 21~cm line & Quantity of H\1\ in the whole region & $\sim 21.4$ \\
 \hline
 \multicolumn{3}{c}{\textbf{Adopted column density for metal abundance derivations}} &  \textbf{$20.75 \pm 0.30$} \\
 \hline
\end{tabular}
\end{center}
\end{table*}

In Table~\ref{tab:hidet} we report the various determinations obtained for the H\1\
column density in NGC604. The neutral hydrogen was also investigated using radio
observations. Israel et al. (1974) noticed a correlation between H\2\ region locations
and large H\1\ complexes in M33. By measuring the H\1\ column density with the 21~cm line
in emission, Dickey et al. (1993) found $\log N\textrm{(H\1)}=21.38\pm0.01$
($45\textrm{"}\times45\textrm{"}$ beam), while Newton (1980) found $21.43\pm0.50$
($47\textrm{"}\times93\textrm{"}$ beam). The direct comparison between these
measurements, which sample the H\1\ in all the region, and the H\1\ detected in
absorption in front of the stellar cluster is not straightforward. Hence we only notice
that the H\1\ quantity measured in emission is larger than the one measured in
absorption, which is what is expected given the source geometry. From now on we adopt the
\textit{FUSE} value for our abundance studies in the neutral medium, i.e., $\log
N\textrm{(H\1)} = 20.75 \pm 0.30$. This value, together with these errors, agrees closely
with all the other determinations.

\section{Heavy elements}\label{sec:heavy}

\begin{figure}
\epsfxsize=7cm \rotatebox{-90}{\epsfbox{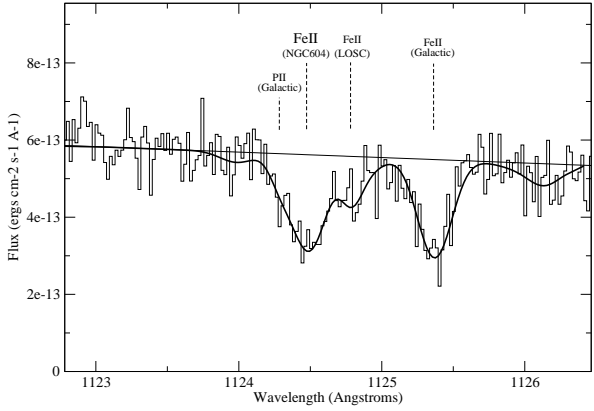}}
  \hspace{0.2cm}
\caption[]{\small{Result of the profile fitting method for the Fe\2\ line at
1125.4478\,\AA. Three absorption components are detected, NGC604, the line of sight cloud
(LOSC) at $\sim -150$\,\kms, and the Galactic component.}} \label{fig:dataFIT}
\vspace{0.1cm}
\end{figure}

We were able to identify many absorption lines in the {\it FUSE} spectra arising from
heavy elements in the neutral ISM of NGC604 (see Fig.~\ref{fig:spectre}). From the
profile fitting of these lines, we derived the turbulent velocities, radial velocities,
and column densities of each species. The complete list of the lines we analyzed is
reported in Table~\ref{tab:lines}. In Fig.~\ref{fig:dataFIT}, we show the example of the
$\lambda1125.4$ Fe\2\ line profile fitting.

\begin{table*}
\caption{\small{Analyzed lines of the metallic species in NGC604, where $f$ is the
oscillator strength, $\lambda_{\rm{rest}}$ the rest wavelength (from the tables of Morton
1991 \&\ 2003), and $\lambda_{\rm{obs}}$ the observed wavelength as obtained with the
independent fit method applied to the LWRS {\it FUSE} spectrum.}}\label{tab:lines}
\begin{center}
\begin{tabular}{lllll}
\hline \hline
 Species & $\lambda_{\rm{rest}}$ & $\lambda_{\rm{obs}}$&  $f$     & comments \\
 \hline
 N\1\   &  963.9903 & 963.218 & 0.124$\times 10^{-1}$ & saturated, blended with Galactic H$_2$ \\
        &  964.6256 &963.853& 0.790$\times 10^{-2} $& not saturated, blended with Galactic H$_2$ \\
        &  965.0413 &964.269& 0.386$\times 10^{-2} $& not saturated, blended with Galactic P\2\ and  N\1\ \\
        &  1134.1653&1133.257& 0.146$\times 10^{-1}$ & not saturated\\
        &  1134.4149&1133.507& 0.287$\times 10^{-1}$ & not saturated\\
        &  1134.9803&1134.072& 0.416$\times 10^{-1}$ & saturated, blended with Galactic  N\1\ \\
 O\1\   & 924.9500  &924.211&  0.154$\times 10^{-2}$ & not saturated, blended with Galactic H$_2$ \\
 &929.5168  &923.778& 0.229$\times 10^{-2}$& barely saturated, blended with Galactic H$_2$ and Ly$\zeta$ \\
 &936.6295  &935.881& 0.365$\times 10^{-2}$& saturated\\
 &971.7382 &970.962&  0.124$\times 10^{-1}$ & strongly saturated, blended with Ly$\gamma$ \\
 &1039.2304 &1038.400& 0.907$\times 10^{-2}$ & strongly saturated, blended with Galactic H$_2$ \\
 Si\2\  &1020.6989&1019.872&0.168$\times 10^{-1}$& not saturated\\
 P\2\   &963.8005&962.997&0.146$\times 10^{1}$ & strongly saturated, blended with Galactic H$_2$ \\
    &1152.8180&1151.857&0.236$\times 10^{0}$& not saturated\\
 Ar\1\  &1048.2198&1047.364&0.263$\times 10^{0}$ & not saturated\\
  &1066.6599&1065.789&0.675$\times 10^{-1}$ & not saturated\\
 Fe\2\  &1055.2617& 1054.457&0.615$\times 10^{-2}$& not saturated \\
&1062.1517&1061.342&0.291$\times 10^{-2}$& barely detected \\
&1063.1764&1062.366&0.600$\times 10^{-1}$& not saturated \\
&1063.9718&1063.160&0.475$\times 10^{-2}$& blended with Galactic Fe\2\ \\
&1081.8748&1081.050&0.126$\times 10^{-1}$& not saturated, blended with Galactic H$_2$ \\
&1096.8770&1096.040&0.327$\times 10^{-1}$& not saturated \\
&1112.0480&1111.200&0.446$\times 10^{-2}$& not saturated \\
&1125.4478&1124.589& 0.156$\times 10^{-1}$ & not saturated \\
&1127.0984&1126.218& 0.282$\times 10^{-2}$ & not saturated \\
&1133.6654&1132.801&0.472$\times 10^{-2}$& not saturated\\
&1142.3656&1141.494&0.401$\times 10^{-2}$& not saturated\\
&1143.2260&1142.354&0.192$\times 10^{-1}$&not saturated, blended with Galactic Fe\2\ \\
&1144.9379&1144.065&0.106$\times 10^{0}$& saturated\\
 \hline
\end{tabular}
\end{center}
\end{table*}

\subsection{Kinematics}\label{sec:cine}

Using the results of the independent-fit method, we were able to compute the turbulent
velocities of each species (see Table~\ref{tab:turb}). For comparison, the turbulence
measured in the ionized gas of the H\2\ region from H$\alpha$ observations is
$28.3\pm1.2$\,\kms~in the core and $20.1\pm2.0$~\kms~in the halo (Melnick 1980). Since
turbulence does not depend on the ionic mass and, in our case, dominates the intrinsic
line broadening (see Sect.~\ref{sec:ext}), we expected {\it a priori} to find similar
values of $b$ for all the species. However, we actually obtain values that are
inconsistent with each other within the errors. This could be due to the fact that for
most species we are only dealing with weak lines whose profile does not depend
significantly on $b$. On the other hand, the small error bars seem to suggest a different
explanation for this inconsistency. The absorption lines we are detecting may be composed
of several unresolved components with various line parameters (i.e., width, column
density, and radial velocity). As a result, the $b$ value derived from a single component
analysis possibly has no physical meaning (Hobbs 1974).

\begin{table}
\caption{\small{Turbulent velocities ($b$) in \kms, neglecting the temperature broadening
component, where errors are given at $2 \sigma$ and 'IF' stands for
 independent fits (see Sect.~\ref{sec:datapfm}).}}\label{tab:turb}
\begin{center}
\begin{tabular}{lll}
\hline \hline
 Species & LWRS, \emph{IF} & MDRS, \emph{IF} \\
 \hline
 N\1\   & $27.9^{+4.7}_{-3.9}$ & $20.1^{+4.8}_{-4.0}$         \\
 O\1\   & $26.5^{+5.3}_{-5.7}$ & $37.2^{+5.4}_{-6.9}$            \\
 Si\2\  & $34.8^{+5.1}_{-7.3}$ & $43.1^{+8.9}_{-6.6}$       \\
 P\2\   & $19.4^{+8.3}_{-3.9}$ & $38.9^{+27.9}_{-27.4}$      \\
 Ar\1\  & $29.7^{+10.5}_{-11.4}$ & $20.4^{+7.1}_{-6.6}$    \\
 Fe\2\  & $33.8^{+2.5}_{-2.2}$ & $56.5^{+4.2}_{-4.2}$    \\
 \hline
 Mean value & $29\pm13$  &   $36\pm14$ \\
 \hline
\end{tabular}
\end{center}
\end{table}

The mean $b$ value derived from the independent fits is similar to the turbulent-velocity
determination from the simultaneous fit, i.e., $29\pm13$ vs. $30.7^{+1.8}_{-1.7}$\,\kms\
for the LWRS spectrum and $36\pm14$ vs. $24.1^{+3.3}_{-3.2}$\,\kms\ for the MDRS
spectrum. This demonstrates that the simultaneous fit effectively averages the $b$
parameter over the different atomic species.

\begin{table}
\caption{\small{Radial velocities of the neutral species absorption lines in \kms, where
'IF' stands for independent fits (see Sect.~\ref{sec:datapfm}) and errors are given at
2~$\sigma$.}}\label{tab:helio}
\begin{center}
\begin{tabular}{lll}
\hline \hline
 Species & LWRS, \emph{IF} & MDRS, \emph{IF} \\
 \hline
 N\1\   & $-242.7^{+3.2}_{-2.8}$ & $-235.7^{+3.8}_{-4.3}$         \\
 O\1\   & $-239.8^{+1.9}_{-1.7}$ & $-231.0^{+3.1}_{-3.6}$            \\
 Si\2\  & $-245.4^{+3.5}_{-2.5}$ & $-245.5^{+4.5}_{-5.9}$       \\
 P\2\   & $-252.4^{+6.2}_{-2.8}$ & $-250.8^{+10.2}_{-12.0}$      \\
 Ar\1\  & $-242.9^{+2.9}_{-2.8}$ & $-246.5^{+3.3}_{-3.1}$    \\
 Fe\2\  & $-239.8^{+1.9}_{-2.7}$ & $-229.2^{+3.8}_{-4.1}$    \\
 \hline
 Mean value & $-244\pm7$ & $-240\pm10$ \\
 \hline
\end{tabular}
\end{center}
\end{table}

In addition to the turbulent velocities, we also derived the radial velocities from the
absorption lines of the neutral species. Given the relatively large number of the lines
we simultaneously analyzed for each species, we were able to reach accuracies smaller
than the instrumental broadening on these determinations. Our results (see
Table~\ref{tab:helio}) suggest that the neutral gas is somewhat less blue-shifted than
the ionized gas in the H\2\ region, for which the radial velocity as inferred from
H$\alpha$ observations is $-256$\,\kms\ (Tenorio-Tagle et al. 2000). A similar trend in
the radial velocities was found in the dwarf irregular galaxy NGC1705 and was attributed
to a different origin of the two gaseous phases within an outflowing superbubble (Heckman
et al. 2001). However, there is no evidence of strong gas ouflows/infall in NGC604.

\subsection{Column densities}\label{sec:metals}

\begin{figure}
\hspace{0.4cm}
 \epsfxsize=8.5cm
\rotatebox{0}{\epsfbox{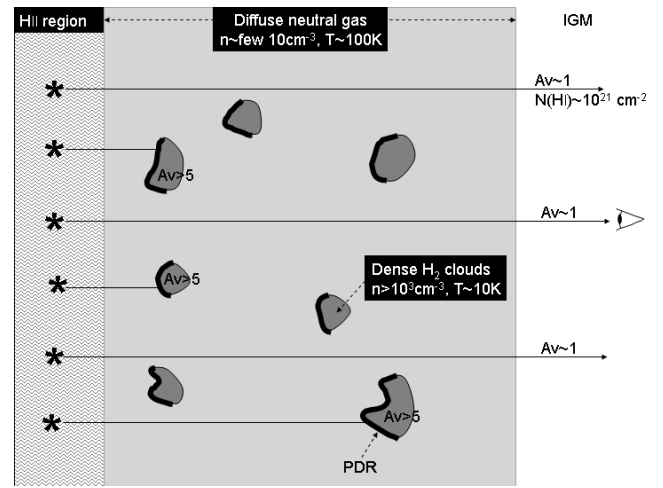}}
  \hspace{0.4cm}
\caption[]{\small{An illustration of the interface between the ionized gas of the H\2\
region and the surrounding neutral medium.}} \label{fig:picture} \vspace{0.1cm}
\end{figure}

One usually models photodissociation regions (PDRs) as a single-interface H\2\
region/molecular cloud, and the material in the PDRs is generally assumed to be
relatively uniform. However, on larger scales, the photodissociation front is likely to
be rather clumpy, composed by small dense molecular clumps embedded in a more diffuse and
mainly neutral medium. The PDRs are located on the surfaces of these clumps (Meixner \&
Tielens 1993). The clump distribution controls the far-UV photon penetration in the sense
that the far-UV radiation passes between the dense clumps and is able to partially ionize
the diffuse gas (see Fig.~\ref{fig:picture}). According to the above-mentioned sketch,
the ISM in NGC604 is probably quite complex and the absorption is likely to arise in
different gaseous phases, i.e., diffuse ionized gas, diffuse neutral gas, or small dense
H$_2$ clouds. However, the densest clouds are opaque to the far-UV radiation and do not
contribute to the total UV spectrum. Moreover, we cannot see the direct contribution of
the PDRs to the absorption spectrum, since the UV radiation generating these PDRs is
blocked by the foreground associated dense molecular clouds. On the other hand, this
contribution could manifest itself in scattered light, since dust is certainly present on
the surfaces of the dense clumps (Ma{\'{\i}}z-Apell\'aniz et al. 2004). However, the
presence of scattered light has not yet been assessed. We thus expect the neutral
absorption lines to arise mainly from the diffuse neutral gas in NGC604, with a possible
contribution from the PDRs.

\begin{table*}
\caption{\small{Metal column densities in NGC604 from the LWRS and MDRS spectra (in
logarithmic units of the column density in \cm), where errors are given at 2~$\sigma$,
'SF' stands for the simultaneous fit method, 'IF' stands for the independent fits (see
Sect.~\ref{sec:datapfm}), and 'LOSC' stands for the line of sight cloud at $\sim
-150$\,\kms.}}\label{tab:CD}
\begin{center}
\begin{tabular}{ccccc|c|c}
\hline \hline
  & NGC604   & NGC604   & NGC604  & NGC604 & LOSC  & Galactic$^{\mathrm{b}}$\\
  & LWRS, \emph{IF} & LWRS, \emph{SF} & MDRS, \emph{SF} & "multi"$^{\mathrm{a}}$ & LWRS, \emph{SF}  &   LWRS, \emph{SF}  \\
 \hline
 N\1\   & $15.31^{+0.34}_{-0.17}$ & $15.14^{+0.02}_{-0.02}$ & $15.26^{+0.06}_{-0.05}$ & $[15.1, \gtrsim16.0]$ & $<12.31^{\mathrm{c}}$ & $15.84$      \\
 O\1\   & $16.52^{+0.19}_{-0.15}$ & $16.26^{+0.05}_{-0.04}$ & $16.40^{+0.34}_{-0.14}$ & $[16.3, \gtrsim17.0]$ & $16.16^{+0.82}_{-0.46}$ & $17.13$    \\
 Si\2\  & $15.54^{+0.11}_{-0.06}$ & $15.52^{+0.04}_{-0.04}$ & $15.62^{+0.10}_{-0.07}$ & $[15.5, \gtrsim16.5]$ & $14.28^{+0.25}_{-0.36}$ & $15.41$   \\
 P\2\   & $13.70^{+0.09}_{-0.08}$ & $13.66^{+0.06}_{-0.07}$ & $13.68^{+0.14}_{-0.15}$ & $[13.7, 14.0]$ & $12.75^{+0.32}_{-0.58}$ & $14.32$     \\
 Ar\1\  & $13.86^{+0.07}_{-0.06}$ & $13.95^{+0.04}_{-0.04}$ & $14.06^{+0.07}_{-0.07}$ & $[13.7, 14.0]$ & $<12.75^{\mathrm{c}}$ & $14.21$   \\
 Fe\2\  & $14.89^{+0.03}_{-0.03}$ & $14.88^{+0.02}_{-0.02}$ & $14.94^{+0.04}_{-0.05}$ & $[14.8, 15.0]$ & $14.51^{+0.15}_{-0.12}$ & $14.95$  \\
 \hline
\end{tabular}
\end{center}
\begin{list}{}{}
\item[$^{\mathrm{a}}$] Results from the multicomponent analysis are discussed in
Sect.~\ref{sec:hidden}. \item[$^{\mathrm{b}}$] Errors on the Galactic estimates are on
the order of $\sim 0.15$ dex. \item[$^{\mathrm{c}}$] Upper limits calculated at 2
$\sigma$.
\end{list}
\end{table*}

In Table~\ref{tab:CD} we report the column densities inferred from the two \textit{FUSE}
observations with both methods, as explained in Sect.~\ref{sec:datapfm}, i.e.,
independent fits (\emph{IF}) for each element and simultaneous fit (\emph{SF}) for
several grouped elements. Determinations of Si\2, P\2, and Fe\2\ column densities, using
the two approaches with the same LWRS spectrum, are consistent within the errors, while
those of N\1, O\1, and Ar\1\ are significantly different. The disagreement could be due
to systematic errors introduced by the \emph{SF} approach. This method assumes that
species share temperatures, heliocentric velocities, and turbulent velocities. The errors
associated to this assumption are not included in the uncertainties we mention.

The \emph{IF} method was not used to determine column densities from the MDRS observation
because of a signa-to-noise ratio too low that results in unstable solutions.
Furthermore, the results from the two different observations (LWRS and MDRS apertures,
\emph{SF} method) do not in general agree with each other well. The inconsistencies could
be explained by the different aperture sizes, together with the large extent of NGC604
cluster in the far-UV (see Sect.~\ref{sec:obs}), or by the systematic effects discussed
in the previous paragraph.

In their study of the diffuse molecular hydrogen content in M33, Bluhm et al. (2003) also
derived O\1, Ar\1, and Fe\2\ column densities in NGC604 (and associated 1$\sigma$ errors)
by using the curve-of-growth method based on the equivalent widths inferred from the same
\textit{FUSE} LWRS spectrum we analyzed. The authors obtained $16.20^{+0.30}_{-0.20}$ for
O\1, which is consistent with our LWRS,~\textit{SF} value, but significantly lower than
the LWRS,~\textit{IF} value. Their derived Ar\1\ column density, $13.65^{+0.15}_{-0.10}$,
is also lower than our determinations. Finally, the estimate of the Fe\2\ column density,
$15.00^{+0.10}_{-0.10}$, is only marginally consistent with our values.

\section{Modelling the ionization structure with CLOUDY}\label{sec:model}

In order to derive abundance ratios from column density measurements, it is generally
assumed that the primary ionization state of one element is representative of its total
abundance in the neutral gas. We expect to find all elements with larger ionization
potential than that of hydrogen (13.6 eV) as neutral atoms in the H\1\ gas. This is the
case for N, O, and Ar, although some fraction of argon (and to a lesser extent nitrogen)
can be singly ionized in low-density neutral regions, due to a large photoionization
cross-section (Sofia \&\ Jenkins 1998). Fe, P, and Si are mostly found as single-charged
ions, with negligible amounts of neutral atoms. Thus N\1, O\1, Si\2, P\2, Ar\1, and Fe\2\
should be the dominant forms of the respective elements in the neutral gas, and their
column densities are thought to be representative of the abundances of the element.

However, ionization corrections may be needed since these atoms or ions can also exist in
the ionized gas of the H\2\ region along the sightlines, contributing to the absorption
lines we observe. In order to estimate this contamination, we modelled the ionized gas
using the photoionization code \texttt{CLOUDY} (Ferland 1996; Ferland et al. 1998). We
assumed that the H\2\ region is a homogeneous Str\"omgren sphere (with a radius $R_s$),
ionized by a single star having the same radiation field as the stellar cluster. Although
this is a very idealized situation, this is certainly sufficient for our purpose of
obtaining rough estimates of the ionization corrections. The input N, O, Ar, and Fe
abundances in the ionized gas are the observed abundances. They are taken from Esteban et
al. (2002), which accounts for electronic temperature fluctuations and provides
consistent abundances of these elements in the ionized gas from the same dataset. The
input P and Si abundances are calculated assuming, respectively, that P/O is equal to the
solar ratio (see Lebouteiller et al. 2005 for P/O measurements in the Milky Way) and that
Si/O is equal to the mean value measured in the ionized gas of BCDs (Izotov et al. 1999).
The hydrogen volumic density is from Melnick et al. (1980). We used two different stellar
continua to constrain our models. In model (1), we use a stellar continuum built upon the
observed flux at all wavelengths. Model (2) simply assumed a Kurucz stellar continuum at
a temperature of 48\,000 K.

\begin{table*}
\caption{\small{Optical emission-line intensities (normalized to $I$(H$\beta$)=100.0)
given by the observations and by our models (see text).}}\label{tab:emlines}
\begin{center}
\begin{tabular}{lccccccccc}
\hline \hline
Lines & H$\gamma$ & H$\alpha$  & $[$O\3$]$  & $[$O\3$]$  & $[$N\2$]$  & $[$N\2$]$  & He\1\  & He\1\  & He\1\  \\
$\lambda$ (\AA) & 4340 & 6563 & 4959 & 5007 & 6548 & 6584 & 4471 & 6678 & 5876 \\
 \hline
NGC604\small{, model (1)}&47.1&291.2&90.3&260.9&11.4&33.6&4.1&3.3&11.7\\
NGC604\small{, model (2)}&47.0&294.0&81.8&236.2&10.0&29.7&4.2&3.4&12.1\\
 \hline
NGC604\small{, Esteban et al. (2002)}&46.6 &291.0 &78.0 &250.0 &9.8 &26.3 &4.4 &3.7 & 13.1 \\
NGC604\small{, Kwitter et al. (1981)}&45.7 &281.8 &67.6 &208.9 &9.3 &28.2 &3.8 &2.7 &8.1 \\
NGC604\small{, Vilchez et al. (1988)}&44.3 &286.0 &77.5 &207.7 &12.4 &33.5 & 3.7 &2.7 &11.5 \\
\hline
NGC5461&46.5 &291.0 & 112.0 &352.0 &10.8 &31.2 &4.4 &3.6 &12.7 \\
NGC5471&47.6 &278.0 &209.0 &640.0 &1.9 &6.2 &4.1 &2.9 &11.8 \\
NGC2363&47.3&278.0 &244.0 &729.0 &0.4 & 1.5 &4.1 &2.9 & 12.3 \\
\hline
\end{tabular}
\end{center}
\end{table*}

\begin{figure}
 \epsfxsize=7.0cm
\rotatebox{-90}{\epsfbox{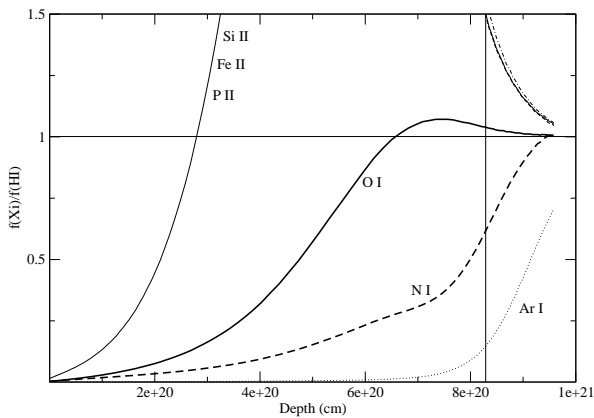}}
  \hspace{0.4cm}
\caption[]{\small{Ionization fraction of each species relative to H\1, as a function of
the distance to the ionizing source. O\1, N\1, and to some extent Ar\1\ are well-coupled
with H\1\ in the neutral gas (i.e., for distances larger than
$R_s=8.20\times10^{20}$\,cm, which is the radius of the Str\"omgren sphere).}}
\label{fig:ionization} \vspace{0.1cm}
\end{figure}

The resulting optical emission-line intensities obtained with both models are given in
Table~\ref{tab:emlines}. For comparison, we report the values of three other giant H\2\
regions. It can be seen that some lines differ significantly for each object, allowing us
to constrain the model. Results of the two models agree reasonably well with each other
and with the observed intensities. From these models we calculated the relative
ionization fractions of each species within the Str\"omgren sphere (see
Fig.~\ref{fig:ionization}) and derived the expected column densities of N\1, O\1, Si\2,
P\2, Ar\1, and Fe\2\ in the ionized gas (assuming that half of the material is in front
of the stellar cluster). These quantities are subtracted from the observed {\it FUSE }
column densities in order to obtain the final column densities in the neutral gas alone
(see column 3 of Table~\ref{tab:ioncorr}). The corrections are relatively small except
for Si\2\ and P\2. This compares well with the findings of Aloisi et al. (2003) in
IZw~18. The authors find that ionization effects are negligible for H, N, O, Ar, and Fe,
the only exception being silicon.

\begin{table*}
\caption{\small{Abundances of N, O, Si, P, Ar, and Fe in the neutral gas, where IC stands
for ionization correction (see Sect.~\ref{sec:model}) and errors are at $2
\sigma$.}}\label{tab:ioncorr}
\begin{center}
\begin{tabular}{lllll|l}
\hline \hline
 &  Ion & IC (dex) & $\log$ (X/H)$^{\mathrm{a}}$ & $[$X/H$]$ & $[$X/H$]_{\rm HII}^{\mathrm{b}}$ \\
\hline
N  & N\1  & $-0.09$ & $-5.70^{+0.30}_{-0.30}$   & $-1.48^{+0.31}_{-0.31}$ & $-0.32^{+0.32}_{-0.32}$ \\
   &      &         & $[-5.7, \gtrsim-4.8]$     & $[-1.5, \gtrsim-0.6]^{\mathrm{c}}$   &    \\
\hline
O  & O\1  & $-0.08$ & $-4.57^{+0.31}_{-0.31}$   & $-1.23^{+0.32}_{-0.32}$ & $+0.00^{+0.22}_{-0.22}$ \\
   &      &         & $[-4.5, \gtrsim-3.8]$     & $[-1.2, \gtrsim-0.5]^{\mathrm{c}}$   &     \\
\hline
Si & Si\2 & $-0.34$ & $-5.57^{+0.31}_{-0.31}$   & $-1.08^{+0.35}_{-0.35}$ & /$^{\mathrm{d}}$ \\
   &      &         & $[-5.6, \gtrsim-4.6]$     & $[-1.2, \gtrsim-0.2]^{\mathrm{c}}$   &                         \\
\hline
P  & P\2  & $-0.15$ & $-7.24^{+0.31}_{-0.32}$   & $-0.60^{+0.33}_{-0.32}$ & /$^{\mathrm{d}}$ \\
   &      &         & $[-7.2, \gtrsim-6.9]$     & $[-0.6, -0.3]^{\mathrm{c}}$          &                         \\
\hline
Ar & Ar\1 & $-0.05$ & $-6.85^{+0.31}_{-0.31}$   & $-1.03^{+0.33}_{-0.33}$ & $+0.25^{+0.20}_{-0.20}$ \\
   &      &         & $[-7.1, \gtrsim-6.8]$     & $[-1.3, -1.0]^{\mathrm{c}}$          &   \\
\hline
Fe & Fe\2 & $-0.04$ & $-5.91^{+0.30}_{-0.30}$   & $-1.36^{+0.31}_{-0.31}$ & $-1.02^{+0.22}_{-0.22}$ \\
   &      &         & $[-6.0, \gtrsim-5.8]$     & $[-1.4, -1.2]^{\mathrm{c}}$          &   \\
 \hline
\end{tabular}
\end{center}
\begin{list}{}{}
\item[$^{\mathrm{a}}$] Values derived from the simultaneous fit method using the {\it
FUSE} LWRS spectrum (see Sect.~\ref{sec:metals}), after respective ionization correction.
\item[$^{\mathrm{b}}$] Ionized gas abundances from Esteban et al. (2002).
\item[$^{\mathrm{c}}$] Values referring to the multicomponent analysis presented in
Table~\ref{tab:CD} and discussed in Sect.~\ref{sec:hidden}.\item[$^{\mathrm{d}}$] No
direct abundance determinations exist in the ionized gas.
\end{list}
\end{table*}

Note that we could not estimate the amount of species in higher ionization stages in the
neutral gas. Indeed Ar\1, and to a less extent N\1, could be partly ionized into Ar\2\
and N\2\ in a low-density ISM, while hydrogen is still into H\1. For this reason, we
might underestimate the argon and nitrogen abundances in the neutral gas when using Ar\1\
and N\1\ for Ar and N. Depending on the hardness of the radiation field, Ar/H can be
larger by $0.2$ up to $0.7$\,dex than Ar\1/H\1\ (Sofia \&\ Jenkins 1998). The situation
is expected to be much less severe for N\1, however.

\section{Chemical abundances in the neutral and ionized gas}\label{sec:ab}

\begin{figure}
 \epsfxsize=7cm
\rotatebox{-90}{\epsfbox{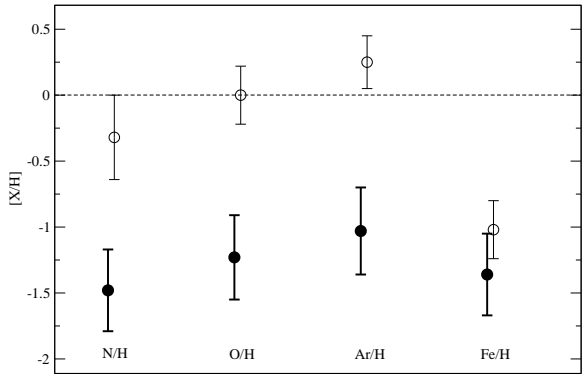}}
  \hspace{0.4cm}
\caption[]{\small{Abundances of N, O, Ar, and Fe in the neutral and ionized gas, compared
to the solar values (from Asplund et al. 2004). We use the notation $[$X/H$]$ = $\log$
(X/H) $-$ $\log$ (X/H)$_\odot$. Empty circles are values for the ionized gas from Esteban
et al. (2002). Filled circles indicate values in the neutral gas as determined from the
simultaneous fit in the \textit{FUSE} LWRS spectrum, using a single-component analysis.
They may be underestimated for N, O, and Ar (see Sect. 9).}} \label{fig:compa1}
\vspace{0.1cm}
\end{figure}

In Table~\ref{tab:ioncorr} we report the abundances toward NGC604 in the neutral gas and,
for comparison, in the ionized gas of the H\2\ region. For the latter we used abundances
given in Esteban et al. (2002), instead of the older values from Kwitter et al. (1981)
and Vilchez et al. (1988). This, because Esteban et al. homogeneously derived the N, O,
Ar, and Fe abundances from a higher resolution optical spectrum, accounting for
electronic temperature fluctuations. We estimate the Si abundance in the ionized gas by
assuming that the abundance ratio Si/O is equal to the mean value measured in BCDs
(Izotov et al. 1999). Abundances are normalized to the solar values from Asplund et al.
(2004) as $[\textrm{X/H}]~=~\log~\textrm{(X/H)}~-~\log~\textrm{(X/H)}_\odot$.
Figure~\ref{fig:compa1} is a graphic representation of these data.

In the neutral gas the abundances of all heavy elements that we obtain are consistent
within $2\sigma$ with the oxygen abundance, which is on the order of $1/10$ solar (see
Table~\ref{tab:ioncorr}). If real (i.e., not driven by the large uncertainties in the
H\1\ column density), this consistency would be somewhat surprising, since we expect at
least Si and Fe to be depleted onto dust grains. Indeed, the medium we are considering
should be comparable to the diffuse neutral medium in front of stars like $\zeta$
Ophiuchi or $\mu$ Colombae in our own Galaxy. N, O, P, and Ar are not depleted or only a
little (within a factor less than $\sim 3$) toward these Galactic sightlines, while Si
and Fe show depletions by a factor of at least $\sim 3$ and 10, respectively (Savage \&\
Sembach 1996; Snow \&\ Witt 1996; Howk, Savage \&\ Fabian 1999). In a similar way, the
heavy element abundances in the ionized gas of the H\2\ region are also consistent within
the errors with the O abundance, which is solar in this case. The only exception is the
gaseous Fe, which is about $1/10$ solar.

If real, the observed abundance trends indicate that nitrogen, oxygen, and argon are
lower by $\gtrsim 1$~dex in the neutral gas phase as compared to the ionized one. This
confirms a similar trend already observed in blue compact dwarf galaxies, and it
contrasts with the most recent finding of no offset in the $\alpha$-element content of
the neutral and ionized gas in the nearby damped Lyman-$\alpha$ galaxy SBS~1543+593
(Bowen et al. 2005). Ionization corrections in the neutral medium cannot fully explain
the similar offsets of N, O, and Ar in NGC604, since Ar is expected to be much more
affected than N and O (Sect.~\ref{sec:model}).

Iron is instead the only element showing similar abundances in the two gaseous phases.
The iron behavior requires that this element has the same gas abundance and depletion
factor in both the neutral and ionized gas phases. This is rather surprising, because we
expect iron to be underabundant in the H\1\ region and depletion on grains to be similar
or larger in the neutral medium compared to the H\2\ region (but see Sect. 9).

\begin{figure}
 \epsfxsize=7cm
\rotatebox{-90}{\epsfbox{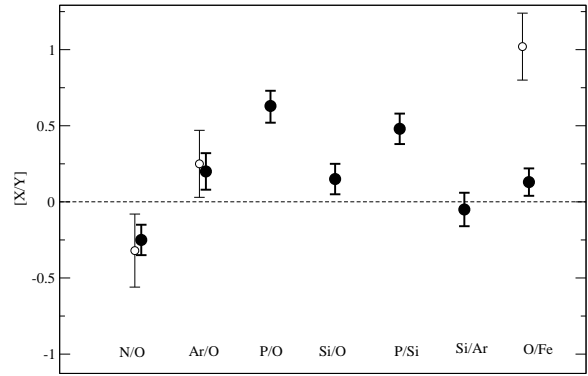}}
  \hspace{0.4cm}
\caption[]{\small{Abundance ratios in the neutral and ionized gas of NGC604. See
Fig.~\ref{fig:compa1} for a description of the plot. The abundances in the neutral gas
weres determined using a single-component model.}} \label{fig:compa2} \vspace{0.1cm}
\end{figure}

For what concerns the relative abundance ratios, we observe the following behaviors (see
Fig.~\ref{fig:compa2}). The $\alpha$-elements O, Si, and Ar are roughly in solar
proportion in both the neutral and the ionized ISM. This seems to indicate that Ar does
not suffer from the ionization effects typical of a low-density neutral medium (see
Sect.~\ref{sec:model}) and that Si is not more depleted in the H\1\ gas than in the H\2\
gas, another rather surprising result. In a similar way, the consistency of the
(subsolar) abundance ratio N/O in both the neutral and ionized gas suggests that N is
affected neither by ionization nor by depletion effects. The behavior of phosphorus in
the neutral gas is also particularly intriguing, since P/O and P/Si in this gas seem to
be super-solar, in disagreement with what found by Lebouteiller et al. (2005): these
authors derived essentially solar P/O ratios for several sightlines sampling the Galactic
diffuse ISM, as well as for a few damped Lyman-$\alpha$ systems and the ISM of the small
Magellanic cloud toward the star Sk108.

We show in the next section that the N\1, O\1, and Si\2\ column density determinations
may be severely affected by saturation of hidden velocity components. If this is indeed
the case, the analysis of Sect. 6, which assumes a single absorbing neutral component
with a large velocity dispersion in front of NGC604, would result in an underestimate of
the abundances of these elements. Once hidden saturation is properly taken into account,
it is possible to easily reconcile all the puzzling results from the behavior of the
various heavy elements in a very simple way, i.e., by considering the possibility that
the neutral and ionized gas could have similar abundances.

\section{Multicomponent analysis}\label{sec:hidden}

Instead of arising from a single neutral cloud with a large velocity dispersion, the
interstellar lines we detect might be the blend of many unresolved absorption components
whose widths and velocities are related to the ISM structure and to the source
morphology. As we now show, this may result in a severe underestimate of the column
densities for the most saturated lines if a single-component analysis is made. It is not
possible with the present data to assess whether multiple saturated components contribute
to the absorption lines we observe. However, it is possible to perform empirical tests
and to calculate the uncertainties associated with the presence of multiple $-$ possibly
saturated $-$ components. A detailed analysis can be found in Lebouteiller (2005). In
this section, we summarize the method and the results.

\subsection{Presence of hidden saturated components}

To identify which lines can suffer from hidden saturation, we first consider the presence
of a test component placed at various velocities within the broad line. We choose for
this component a $b$-value of $2$\,\kms, which we consider as the minimum value allowed
for a single sightline intersecting a single interstellar cloud (Tumlinson et al.~2002
measured $b$ values systematically larger than $1.6$\,\kms\ toward stars of the
Magellanic Clouds).

Among the detected lines in NGC604, Fe\2\ $\lambda1142$ is the only one for which the
test component can never be saturated due to the relatively low value of the oscillator
strength. The other lines of Fe\2\ and all the lines of the other species could suffer
from hidden saturation by a component with $b$ = $2$\,\kms. Hence, the column density
derived from the observed global $\lambda1142$ Fe\2\ line should not be affected by the
systematic errors associated with hidden saturated components, even in the center of the
global line, where the optical depth is maximum. Note that it can still suffer from
uncertainties due to the non-linear sum of the individual spectra (see previous section).
The  Fe\2\ column density obtained using the $\lambda1142$ line alone is $\log
N(\textrm{Fe\2})=14.97\pm0.07$, slightly higher than the value we found earlier using all
the available lines ($14.88^{+0.02}_{-0.02}$, see Sect.~\ref{sec:metals}). This confirms
that using stronger Fe\2\ lines in addition to $\lambda1142$ introduces systematic errors
if there are hidden saturated components. However, the errors do not exceed
$\sim0.1$\,dex in the case of Fe\2.

For other species, we cannot rule out that even their weakest observable lines could be
made of saturated individual components. Although we have no information on the actual
velocity component distribution responsible for the global profile, it is still possible
to test various plausible distributions, which can somehow be constrained if we
simultaneously adjust and fit all the lines of a given species, from the weakest line to
the heavily saturated one.

\subsection{Test of possible component distributions}

We consider here various component distributions, fully defined by the number of
components $n_c$, by the turbulent velocity $b$ of each component, and by the spacing
${\Delta}V$ between them (which can possibly be lower than $b$, depending on the number
of components).

The ideal species to test for plausible distributions is Fe\2\ because of the large
number of lines available, with a wide range of oscillator strengths. Fe\2\ might have a
space distribution differing slightly from that of the other species, due to abundance
and ionization inhomogeneities. However, here we only wish to test the method, so we
ignore this problem.

We consider various distributions with $n_c$ varying between 1 and 20 and $b$ between 2
and 44\,\kms\ (which is the highest $b$ value for a single component, see
Sect.~\ref{sec:cine}), while ${\Delta}V$ is set so that the components are uniformly
distributed within the observed Fe\2\ line widths. For a given line, the column densities
in each component are considered as free parameters by the fitting procedure
\texttt{Owens}.

We find that many component distributions provide a $\chi^2$ equal to or lower than the
$\chi^2$ for a single-absorption component. We thus consider all these distributions as
mathematically and physically plausible. The distributions give a total column density
(sum of all the components) in agreement within 0.1\,dex with the determination using a
single component. Furthermore, we notice that satisfactory distributions implying
components with low $b$ values (potentially responsible for saturation effects) also
imply a large number of components $n_c$, so that the total column density is dispersed
in many components with relatively low column densities. Typically, at least 13
components with $b=2$\,\kms\ were needed to adjust the lines. We do not find satisfactory
distributions implying high column densities together with low $b$-values.

We used the velocity distributions that we found satisfactory to fit the Fe\2\ lines, to
also fit the lines of the other species (N\1, O\1, Si\2, P\2, and Ar\1). We discovered
that the distributions do not introduce strongly saturated components in the P\2\ and
Ar\1\ lines, even when $b=2$\,\kms. The column density we determined using the velocity
distributions is similar to the estimate using a single-component fit within
$\sim0.2$\,dex. However, several possible distributions imply that even the weakest N\1,
O\1, and Si\2\ lines are made of saturated components. For these species, the sum of the
column densities in the individual components can be larger than $2-3$\,dex in comparison
with the result of the single-component analysis.

Thus the column densities of N\1, O\1, and Si\2\ could be severely underestimated. On the
other hand, the Fe\2\ and, to a lesser extent, the P\2\ and Ar\1\ lines do not seem to be
strongly affected by systematic errors due to the presence of hidden saturated
components. We show the ranges of column densities derived from the multicomponent
analysis in Table~\ref{tab:CD}.

Focussing on the P\2, Ar\1, and Fe\2\ column densities only, the interpretation of the
abundances in the neutral gas differs from what is discussed in Sect.~\ref{sec:ab}. Iron
has similar abundances in the ionized and neutral phases, suggesting that the two gaseous
phases could indeed have similar metallicities. The underabundance of Ar\1\ could then be
explained by the fact that argon can be partly ionized into Ar\2\ in this gaseous phase.
The actual Ar/H would be closer to Ar\2/H\1. This result would agree with the findings in
IZw36 (Lebouteiller et al. 2004). Finally, little can be said about the abundances of
N\1, O\1, and Si\2, which could well be similar to those in the H\2\ region. To conclude,
the evidence of a difference in abundances in the H\1\ and in the H\2\ gases essentially
vanishes as a result of our tests.

\section{Conclusion}

This study provides the first detailed analysis of interstellar lines of H\1, N\1, O\1,
Si\2, P\2, Ar\1, and Fe\2\ in the neutral medium in front of a giant H\2\ region in the
spiral galaxy M33. Since NGC604 is a nearby system, we have been able to perform the
necessary critical tests for analyzing possible selection effects.

\begin{itemize}

\item  In the frame of a simple model, we have derived column densities of metals in
    the neutral gas
    of NGC604 from both \textit{FUSE} LWRS and MDRS spectra. These
    independent estimates allowed us to quantify the effects of
    the source extent on the interstellar absorption line profiles.

\item The continuum used for the profile fitting
    was checked by comparison with a theoretical stellar model.
    Besides, no significant contamination from stellar photospheric lines was found for
    the H\1 absorption lines we investigated.

\item A particular attention was given to the neutral hydrogen
    column-density determination. The Ly$\beta$ line from the \textit{FUSE}
    LWRS and MDRS spectra is contaminated by the O\6\ P~Cygni doublet. We modelled
    this contamination to obtain the H\1\ column density. Also, profile fitting
    of the Ly$\alpha$ absorption in HST/\textit{STIS} spectra toward individual
    stars in the cluster NGC604 reveals inhomogeneities in the neutral gas. We
    finally adopted the FUSE value of $\log~N$(H\1)=20.75 with a conservative error
    of $\pm$0.3~dex to account for all the possible uncertainties. Within the errors,
    this H\1\ column density is consistent with the \textit{IUE}
    determination using Ly$\alpha$ and with the 21~cm line radio observations.

\item By modelling the ionization structure of the H\2\ gas with the
    photoionization code \texttt{CLOUDY}, we have shown that N\1, O\1, Ar\1,
    and Fe\2\ are reliable tracers of the neutral gas, in contrast with Si\2\
    and P\2, which require ionization corrections to obtain final abundances
    in the neutral phase.

\end{itemize}

Adjusting absorption lines with a single component, we find that N, O, Ar, and Si are
underabundant in the neutral gas as compared to the ionized gas by factors $\gtrsim10$,
while Fe/H is similar in the two gaseous phases. This result is rather puzzling, since
iron is expected to be equally or more depleted in grains in the neutral gas compared to
the ionized one, and there is no reason it should be relatively more abundant in the
neutral gas.

This led us to investigate in detail the influence of individual unresolved components in
the analysis of absorption lines. Using the method and results of Lebouteiller (2005), we
argue that N\1, O\1, and Si\2\ column densities can be severely underestimated if there
are saturated hidden components, while Fe\2\ and, to some extent, Ar\1\ and P\2\ should
be more reliably determined. Since Fe is the only element to show similar abundances in
the neutral and ionized gas, it is possible that all elements indeed have similar
abundances in both media. The underabundance of Ar would then be due to the fact that we
used Ar\1/H\1\ to estimate Ar/H, when (Ar\1+Ar\2)/H\1\ should be used instead.

\begin{acknowledgements}
This work is based on data obtained by the NASA-CNES-CSA \textit{FUSE} mission operated by
the Johns Hopkins University.
This work used the profile fitting procedure \texttt{Owens.f}
 developed by M. Lemoine and the French \textit{FUSE} Team.
      V.L. is grateful for the
    hospitality of STScI where part of this work was done. We thank Claus
    Leitherer, Ken Sembach, Tim Heckman, Ron Allen, and Jeff Kruk for useful
    discussions and F. Bruhweiler and C. Miskey for having kindly provided HST/\textit{STIS}
    individual spectra. The authors would like to thank the referee for the useful
    comments.

\end{acknowledgements}

\end{document}